\documentclass[compsoc, conference, a4paper, 10pt, times]{IEEEtran}
\usepackage{amsmath, amssymb, amsfonts}
\usepackage[numbers]{natbib}
\usepackage{pgfplots}
\usepackage{hyperref}
\usepackage{comment}
\usepackage{blindtext}
\usepackage{booktabs, multirow}
\usepackage{graphicx}
\usepackage{algorithm, algpseudocode}
\usepackage{siunitx}
\usepackage{times}
\usepackage[
colorinlistoftodos,
textsize=tiny
]{todonotes}
\usepackage{soul}
\usepackage[font=footnotesize]{subcaption}
\usepackage[normalem]{ulem}
\usepackage{titlesec}
\titleformat{\paragraph}[runin]
{\normalfont\normalsize\bfseries}{\theparagraph}{1em}{}[.]
\titlespacing*{\paragraph}{\parindent}{0pt}{1em}

\newcommand{\A}{$\mathbb{A}$\xspace}
\newcommand{\B}{$\mathbb{B}$\xspace}
\newcommand{\DSA}{${DS}_\mathbb{A}$\xspace}
\newcommand{\DSB}{${DS}_\mathbb{B}$\xspace}
\newcommand{\pu}{property unlearning\xspace}
\newcommand{\puBegin}{Property unlearning\xspace}
\newcommand{\expMnist}{$\mathcal{E}_\textit{MNIST}$\xspace}
\newcommand{\expCensus}{$\mathcal{E}_\textit{Census}$\xspace}
\newcommand{\expUtk}{$\mathcal{E}_\textit{UTK}$\xspace}
\newcommand{\expMnistPre}{$\mathcal{E}_\textit{MNIST}^\textit{pre}$\xspace}
\newcommand{\expCensusPre}{$\mathcal{E}_\textit{Census}^\textit{pre}$\xspace}
\newcommand{\expUtkPre}{$\mathcal{E}_\textit{UTK}^\textit{pre}$\xspace}
\newcommand{\expMultiAR}{$\mathcal{E}_\textit{multi}^\text{age\textrangle race}$\xspace}
\newcommand{\expMultiRA}{$\mathcal{E}_\textit{multi}^\text{race\textrangle age}$\xspace}
\newcommand{\targetM}{$\mathcal{M}$\xspace}
\newcommand{\advA}{$\mathcal{A}$\xspace}
\newcommand{\advAone}{$\mathcal{A}_1$\xspace}
\newcommand{\advAtwo}{$\mathcal{A}_2$\xspace}
\DeclareMathAlphabet{\mathpzc}{OT1}{pzc}{m}{it}

\newlength{\figwidth}
\newlength{\figheight}
\begin{document}
	
	\title{Lessons Learned: Defending Against Property Inference Attacks}
	% \titlerunning{Property Unlearning: A Defense Strategy Against PIAs}
	
	% IEEE author block
	 \author{
		 	\IEEEauthorblockN{Joshua Stock\IEEEauthorrefmark{1}, Jens Wettlaufer\IEEEauthorrefmark{2}, Daniel Demmler\IEEEauthorrefmark{1}, Hannes Federrath\IEEEauthorrefmark{1}}
			\IEEEauthorblockA{\IEEEauthorrefmark{1}Universität Hamburg, 		Email: firstname.lastname@uni-hamburg.de \\
				\IEEEauthorrefmark{2} wettlaufer@ieee.org
			}
		 }

	% LNCS author block
	% First Name, Last Name\inst{1}\orcidID{2222-3333-4444-5555} \and
	%  \author{Joshua Stock\inst{1}\orcidID{0000-1111-2222-3333} \and
		%  Jens Wettlaufer \and
		%  Daniel Demmler\inst{1}\orcidID{0000-0001-6334-6277} \and
		%  Hannes Federrath\inst{1}\orcidID{3333-4444-5555-6666}}
	%  %
	%  \authorrunning{J. Stock et al.}
	%  \institute{Universität Hamburg, Germany\\
		%  \email{\{firstname.lastname\}@uni-hamburg.de}}
	
	\maketitle
	
	\begin{abstract}
		This work investigates and evaluates multiple defense strategies against \emph{property inference attacks} (PIAs), a privacy attack against machine learning models.
		Given a trained machine learning model, PIAs aim to extract statistical properties of its underlying training data, e.g., reveal the ratio of men and women in a medical training data set.
		While for other privacy attacks like membership inference, a lot of research on defense mechanisms has been published, this is the first work focusing on defending against PIAs.
				
		With the primary goal of developing a generic mitigation strategy against white-box PIAs, we propose the novel approach \emph{\pu}.
		Extensive experiments with \pu show that while it is very effective when defending target models against specific adversaries, \pu is not able to generalize, i.e., protect against a whole class of PIAs.
		To investigate the reasons behind this limitation, we present the results of experiments with the explainable AI tool LIME.
		They show how state-of-the-art property inference adversaries with the same objective focus on different parts of the target model.
		We further elaborate on this with a follow-up experiment, in which we use the visualization technique t-SNE to exhibit how severely statistical training data properties are manifested in machine learning models.
		Based on this, we develop the conjecture that post-training techniques like \pu might not suffice to provide the desirable generic protection against PIAs.
		As an alternative, we investigate the effects of simpler training data preprocessing methods like adding Gaussian noise to images of a training data set on the success rate of PIAs.
		We conclude with a discussion of the different defense approaches, summarize the lessons learned and provide directions for future work.
	\end{abstract}

	
	\section{Introduction}
	The term \emph{machine learning} describes a class of self-adapting algorithms which fit their behavior to initially presented training data.
	It has become a very popular approach to model, classify and recognize complex data such as images, speech and text.
	Due to the high availability of cheap computing power even in smartphones and embedded devices, the presence of machine learning algorithms has become a common sight in many real-world applications.
	At the same time, issues related to privacy, security, and fairness in machine learning are increasingly raised and investigated.
	
	This work focuses on machine learning with artificial neural networks~(ANNs).
	After an ANN has been constructed, it can ``learn'' a specific task by processing big amounts of data in an initial training phase.
	During training, the connections between the network's nodes (or \emph{neurons}) are modified such that the performance of the network regarding the specified task increases.
	After a successful training phase, the model, i.e., the network, is able to generalize, and thus enables precise predictions even for previously unseen data records.
	But while the model needs to extract meaningful properties from the training data to become good in its dedicated task, it usually ``remembers'' more information than it needs to~\cite{zhang_understanding_2021,song2017machine}.
	This can be particularly problematic if training data contains private and sensitive information such as intellectual property or health data.
	The unwanted manifestation of such information, coupled with the possibility to retrieve it, is called \emph{privacy leakage}.
	In recent years, a new line of research has evolved around privacy leakage in machine learning models, which investigates privacy attacks and possible defense mechanisms~\cite{rigaki2020survey}.
	
	In this paper, we focus on a specific privacy attack on machine learning models: the \emph{property inference attack}~(PIA), sometimes also referred to as \emph{distribution inference} attack~\cite{ateniese2015hacking,ganju2018property}.
	Given a trained machine learning model, PIAs aim at extracting statistical properties of its underlying training data set.
	The disclosure of such information may be unintended and thus dangerous as the following example scenarios show:
	\begin{enumerate}
		\item Computer networks of critical infrastructures have collaboratively trained a model on host data to detect anomalies.
		In this case, a PIA could reveal the distribution of host types in the network to refine a malware attack.
		\item Similarly, a model within a dating app has been trained on user data to predict good matches.
		Another competing dating app could use a PIA to disclose properties of the customer data to improve its service, e.g., the age distribution, to target advertisements more precisely.
	\end{enumerate}
	If such models are published or leaked to the public on other channels, PIAs can reveal secrets of their training data.
	These secrets do not need to be in obvious correlation to the actual model task, like the property \emph{host type} in the anomaly detection model of example 1.	
	
	\subsection{Contributions}
	To the best of our knowledge, no dedicated \textbf{defense mechanism against property inference attacks} (PIAs) has been proposed yet.
	Our primary goal is to harden readily trained ANNs, further called \emph{target models}, against such an adversarial extraction of one or more predefined properties in their training data sets.
	The defense technique should allow to deliberately prune chosen statistical training data properties from a trained model, while keeping its utility as high as possible, thus protecting the privacy of the data set used for training.
	
	We thus present \emph{\pu}, a strategy designed for the \textbf{strong white-box attack scenario}, where the adversary has full access to all internal parameters of the target model, which are learned during the training phase\footnote{In contrast, a weaker black-box adversary only has oracle access to the target model, i.e., is able to solely observe the output of the model for chosen inputs, but does not know internal parameters of the model.}.	
	We have conducted \textbf{thorough experiments} which show, that (a) \pu allows to harden ANNs against a specific PI attacker with small utility loss, but (b) it is not possible to use our approach to completely prune a property from a trained model, i.e., to defend against \emph{all} PI attackers for a chosen property in a generic way.
	
	Consequently, we have conducted further experiments with the \textbf{explainable AI} tool LIME~\cite{ribeiro2016should} and the visualization framework t-SNE~\cite{van2008visualizing}.
	Both provide evidence for the conjecture that properties are deeply rooted and ubiquitous in the trained weights of an ANN, such that complete pruning of a property from a trained ANN is not possible without greatly limiting its utility.
	
	In addition, we investigate the impact of simple training data preprocessing steps such as adding Gaussian noise to images of a training data set on the success rate of PIAs.
	This is meant as an inspiration for possible alternatives to techniques such as differential privacy, which has been established as a de-facto standard against many privacy attacks with the exception of PIAs~\cite{rigaki2020survey, suri2021formalizing, suri2022subject}.
	
	A comprehensive summary of our \textbf{key findings} is provided in Section~\ref{sec:lessons-learned}.
	
	\subsection{Organization of this paper}
	The remainder of this paper is organized as follows:
	Section~\ref{sec:background} briefly explains artificial neural networks, machine learning privacy attacks, and introduces our notation.
	We conclude this section with a description of our threat model and a detailed explanation of property inference attacks.
	Section~\ref{sec:relwork} deals with an overview of related work.
	Our proposed defense strategy \pu is presented in Section~\ref{sec:pu}.
	Section~\ref{sec:pu-exp} describes our \pu experiments, including our findings regarding its limited defense capabilities.
	In Section~\ref{sec:lime}, we experimentally explore the reasons behind the limitations of our defense strategy via the explainable AI tool LIME.
	We develop the conjecture that properties are deep-seated in the trained weights of ANNs and provide experimental evidence with t-SNE visualization in Section~\ref{sec:t-sne}.
	As an additional experiment, we investigate the impact of simple training data preprocessing techniques such as adding Gaussian noise on images, see Section~\ref{sec:preprocessing}.
	We summarize and discuss our findings as well as provide directions for future work in Section~\ref{sec:discussion}.
	Section~\ref{sec:conclusion} concludes this paper.

	\section{Background}
	\label{sec:background}
	In this section, we define our notation and provide an introduction into the fundamental topics of ANNs and privacy attacks against machine learning schemes.
	We also define our threat model and explain PIAs and their background in more detail.
	
	\subsection{Notation}
	\label{sec:notation}
	We denote the set of integers $[k] = \{1,\ldots,k\}$.
	Properties of a data set are denoted as blackboard bold, e.g., \A and \B.
	Replacing the property-subscript with an $*$, we reference all possible data sets $\textit{DS}$, e.g., $\textit{DS}_*$ means both $\textit{DS}_\mathbb{A}$ and $\textit{DS}_\mathbb{B}$.
	An \emph{absolute} increase of $x$~percent points is denoted as $+x\%$P.

	\subsection{Artificial Neural Networks}
	\label{sec:ML-basics}
	An artificial neural network~(ANN) consists of interconnected neurons, organized in multiple layers. Inputs are propagated through the network layer by layer. 
	For this, each neuron has an associated \emph{weight} factor $w$ and a \emph{bias} term $b$. A (usually non-linear) activation function $\sigma$ computes each neuron's output on a given input, specifically for a neuron $n$ and input $x$:
	\(
	n = \sigma(w\cdot x+b)
	\)
	
	Prior to training an ANN, all neurons are individually initialized with random weights and biases (also called \emph{parameters}).
	Utilizing a labeled training data set in an iterative \emph{training} process, e.g., batch-wise backpropagation, these parameters are tuned such that the network predicts the associated label to its given input.
	The speed of this tuning process, respectively its magnitude per iteration, is controlled by the \emph{learning rate}.
	The higher the learning rate, the more the parameters are adapted in each round.
	
	For example, the task of the MNIST data set~\cite{lecun1998gradient} aims at classifying the correct digit of given handwritten digits from 0--9.
	Before training, the neural network with initially random parameters might misclassify a large part of the given examples, correctly identifying coincidentally a few digits (corresponding to a low \emph{accuracy}).
	After the training process, the ANN might have a high accuracy, typically close to 100\%~\cite{lecun1998gradient}, classifying almost all of the given examples correctly and thus being reliable.

	\subsection{Machine Learning Privacy Attacks}
	In general, privacy attacks against machine learning models extract information about training data of a target model \targetM or the target model itself from its trained parameters.
	Some attacks, like membership inference~\cite{shokri2017membership} extract information about a single record from a machine learning model.
	Other attacks try to recover the model itself (model extraction~\cite{papernot2017practical}) or to recover the training data set or parts of it (model inversion~\cite{fredrikson2015model}).
	In contrast, this paper focuses on \emph{property inference attacks}~(PIAs), which reveal statistical properties of the entire training data set.
	This is not to be confused with \emph{attribute inference attacks}, e.g.,~\cite{song2019overlearning}, which enable the adversarial recovery of sensitive attributes for \emph{individual} data records from the training data set.

	%Possible main source: \cite{jegorova_survey_2021}
	
	\subsection{Threat model}
	%White-box motivated in other paper (Eternal Sunshine):
	%``Since at this point it is unclear how much information about a model can be recovered by looking only at its inputs and outputs, to avoid unforeseen weaknesses we characterize forgetting for the stronger case of white-box attacks, and derive bounds and defense mechanism for it.''\\
	In the remainder of this paper, the following threat model is assumed:
	A model owner has trained and shared the model of a neural network.
	The owner wishes to keep their training data and its property \A\ or \B~(a statistical property of the training data) secret.
	An example may be a company that has trained a model on its customer data and does not want to disclose any demographic information about their customers.
	If an attacker gets access to this model, they can perform a PIA and reconstruct the demographics of its training data, breaching the desired privacy.
	In another scenario, an attacker might want to gather information about a computer network before launching a malware attack.
	Such networks are often monitored by intrusion detection systems~(IDS), which have been trained on the network traffic to detect unusual behavior (indicating an intrusion).
	Having access to this IDS model, the attacker could infer the operating system most computers are running on in the system, or even detect specific vulnerabilities by checking the presence of security patches in the network (as demonstrated in~\cite{ganju2018property}).
	
	Furthermore, we assume that the attacker has full \emph{white-box access} to the target model \targetM.
	This means that the attacker can access all parameters and some of the hyperparameters of \targetM:
	The adversary has a complete overview of the ANN architecture and can access the values of the weights and biases of all neurons.
	The adversary also has access to other useful hyperparameters of \targetM such as the batch size during training, the utilized learning rate and the number of training epochs.
	This helps the adversary to tailor their shadow models (see the following Section~\ref{sec:pia}) as close to the target model as possible.
	
	In contrast, an adversary in a \emph{black-box scenario} typically has oracle-access to the target ML model \targetM, allowing only to send queries to \targetM and to analyze the corresponding results, i.e., the classification of a data instance.
	Many machine learning as a service (MLaaS) products are examples for real-world black-box scenarios, where customers can pay to get oracle access to potential target models.
	Note that our approach protects against the \emph{stronger} white-box adversary.
	
	As assumed in previous defenses against ML privacy attacks~\cite{nasr_machine_2018, song2021systematic, tang2021mitigating}, the attacker can access parts of the target model's training data, or knows a distribution of the training data, but does not know the whole training data set itself.
	Alternatively, information about the training data may also be reconstructed like in~\cite[p.~5]{shokri2017membership}, which is just as effective for privacy attacks~\cite{liu2021ml}.
	
	\subsection{Property Inference Attacks (PIAs)}
	\label{sec:pia}
	
	%\todo[inline]{JS: Motivation PIA violates intellectual property, PIA can serve as building block for membership inference (aus PIA-GAN Paper), hier oder woanders}
	
	\citeauthor{ateniese2015hacking}~\cite{ateniese2015hacking} were the first to introduce property inference attacks~(PIAs), with a focus on hidden markov models and support vector machines.
	In this paper, we refer to the state-of-the-art PIA approach by \citeauthor{ganju2018property}~\cite{ganju2018property} who have adapted the attack to fully connected neural networks~(\mbox{FCNNs}), a popular sub-type of artificial neural networks (ANN).
	In a typical PIA scenario, an adversary has access to a trained machine learning model called \emph{target model} \targetM, but not its training data.
	By using the model at inference time, a PIA enables the adversary to deduce information about the training data which the model has learned.
	Since the adversary's tool for the attack is a machine learning model itself, we call it \emph{adversarial meta classifier} \advA. 
	Thus, the adversary attacks the target model \targetM by utilizing \advA to extract a property from its training data.
	
\begin{figure}[]
		\centering
		\includegraphics[width=1.0\linewidth]{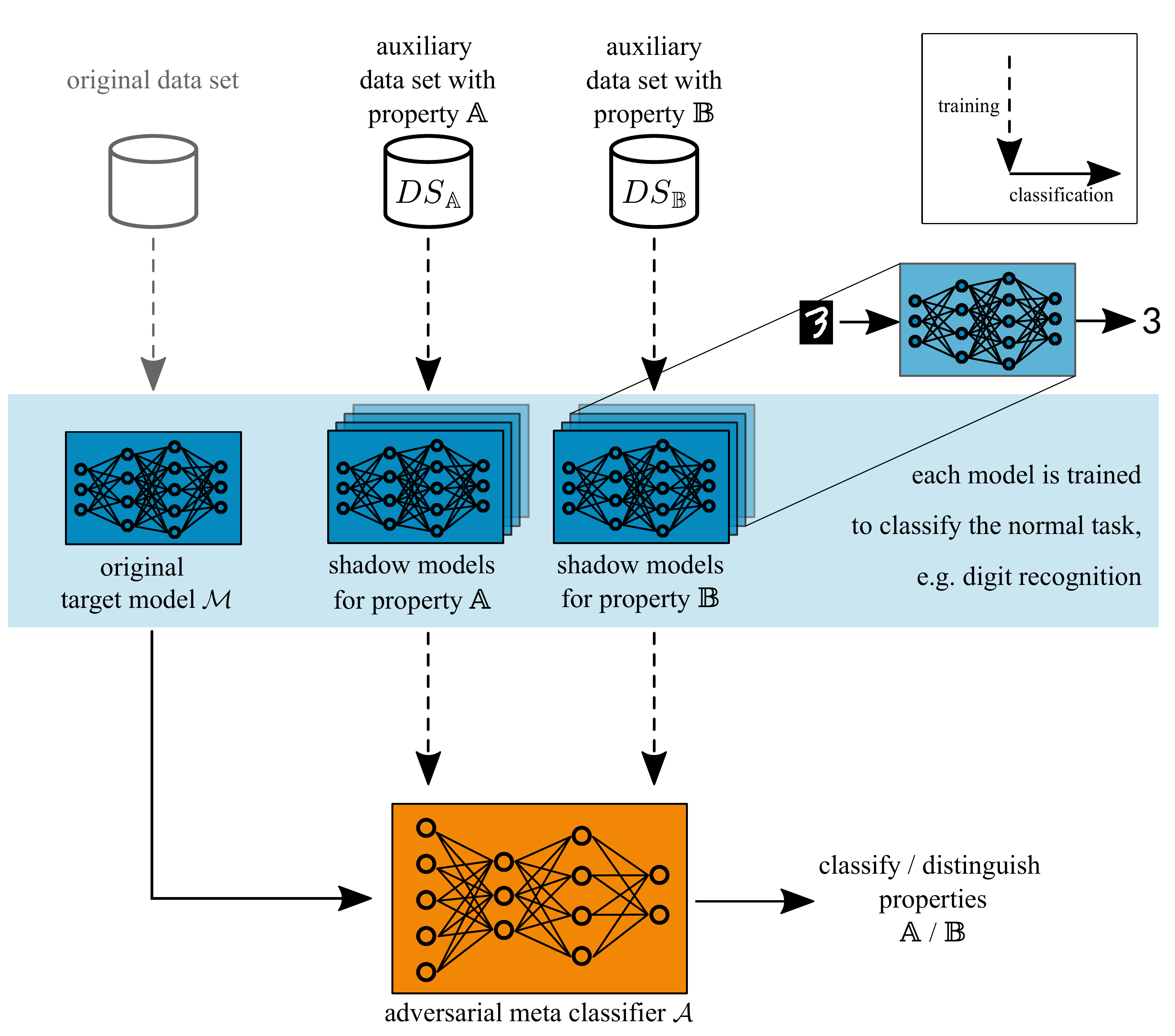}
		\caption{Property inference attack (PIA). The attacker does not have access to the original data set, but to the target model \targetM. If the attacker targets property \A, \DSA and ${DS}_{\overline{\mathbb{A}}}$, i.e., \DSB, need to be created. Then, multiple shadow models are trained based on the created data sets with the same task as \targetM, such as digit recognition. After that, the adversarial meta classifier \advA will train on the shadow models in a supervised manner, until \advA is capable to infer the training data property \A or \B from \targetM.}
		\label{fig:pia}
	\end{figure}

	A PIA typically involves the following steps~\cite{ganju2018property}, as depicted in Figure \ref{fig:pia}:
	\begin{enumerate}
		\item Define (at least) two global properties about the target model's training data set, for example \A\ and \B.
		A successful PIA will show which property is true or more likely for the training data set of the given target model.
		\item For each defined property, create an \emph{auxiliary data set} $DS_*$, i.e., \DSA and \DSB. Each auxiliary data set fulfills the respective property.
		\item Train multiple \emph{shadow models} on each auxiliary data set $DS_*$. \emph{Shadow models} have the same architecture as the target model.
		Due to the randomized nature of ML training algorithms the parameters, i.e., weights and biases, of every model have different initial values.
		\item After training the shadow models, use their resulting parameters (weights and biases) to train the adversarial meta classifier \advA.
		During this training, the meta classifier \advA learns to distinguish the parameters of target models that have been trained on data sets with property \A and data sets with property \B, respectively.
		As a result, \advA\xspace is able to determine which of the properties \A or \B is more likely to be true for the training data of a given target model.
	\end{enumerate}
	
	As an example, suppose the task of a target model~\targetM is smile prediction with \num{50000} pictures of people with different facial expressions in its training data set.
	For a PIA, the adversary defines two properties \A\ and \B\ about the target model's training data set, e.g.,
	\begin{align*}
		\mathbb{A}\,\text{:}\, & \text{proportion of male:female data instances is 0.7:0.3}\\
		\mathbb{B}\,\text{:}\, & \text{male and female instances are equally present}
	\end{align*}
	Given~\targetM, the task of the adversary is to decide which property describes \targetM's training data set more accurately.
	As mentioned in step 2), the adversary first needs to create two auxiliary data sets $DS_\mathbb{A}$ and $DS_\mathbb{B}$, with the male:female ratios as described in the properties above.
	After training shadow models on the auxiliary data sets, the adversary uses the trained weights and biases of the shadow models to train the adversarial meta classifier \advA, which is ready for the adversarial task after its training.\\
	An extension of the meta classifier to more than two properties is trivial:
	For $k$ properties, the adversary needs $k$ auxiliary training data sets, trains shadow models in $k$ groups and constructs \advA as a classifier with $k$ outputs instead of two.
	
	\subsubsection{Formalization}
	The above setup of a property inference attack is covered by the approach of Suri and Evans~\cite{suri2021formalizing} to formalize property inference attacks.
	For their definition, they introduce the publicly accessible common distribution $\mathpzc{D} = (\text{data}~\mathpzc{X}, \text{labels}~\mathpzc{Y})$, the model trainer $\mathpzc{T}$, the adversary $\mathpzc{A}$ and two public functions $\mathpzc{G_0}$ and $\mathpzc{G_1}$ that are able to transform distributions. 
	On the basis of a cryptographic game definition inspired by Yeom et al.~\cite{yeom2018privacy}, they formalize PIAs as follows.
	First, $\mathpzc{T}$ randomly chooses one of the distribution functions $\mathpzc{G_0}$ or $\mathpzc{G_1}$. The resulting distribution is used to sample a data set $\mathpzc{S}$. Then, $\mathpzc{T}$ trains a model $\mathpzc{M}$ on data set $\mathpzc{S}$, which is transferred to the attacker $\mathpzc{A}$. Now, $\mathpzc{A}$ has the goal to infer the initially chosen distribution transformer function $\mathpzc{G_0}$ or $\mathpzc{G_1}$.
	Thus, the adversary aims at distinguishing the sub-distributions, $\mathpzc{G_0(D)}$ or $\mathpzc{G_1(D)}$, which reflect the training distribution of $\mathpzc{M}$. Since $\mathpzc{G_0}$ and $\mathpzc{G_1}$ are defined by the attacker, the correct hypotheses leads to a successful attack.
	For further details and limitations of the formalization approach, please refer to~\cite{suri2021formalizing}.

	\subsubsection{Permutation Invariance}
	The two types of ANNs we consider are fully connected neural networks~(FCNNs) and convolutional neural networks~(CNNs).
	Both have the property that the neurons within each ``dense''-layer are interchangeable, since each neuron of one layer is equally connected to all neurons of the previous and the following layer.
	Therefore, the ordering of neurons within a \mbox{dense-layer} is insignificant.
	
	However, if the neuron parameters (weights and biases) are processed as they are by an adversarial meta classifier $\mathcal{A}$, i.e., in their given order, $\mathcal{A}$ has to find patterns in differently ordered weights and biases.
	This makes it inherently more difficult for the meta classifier to learn the patterns of the shadow model weights successfully, because each shadow model is trained individually and most likely has its neurons ordered differently.
	
	As a solution, \citeauthor{ganju2018property} have explored the possibility of processing the neuron parameters of each ANN-layer \emph{as a set}.
	This permutation invariant approach, inspired by \citeauthor{zaheer2017deep}~\cite{zaheer2017deep}, eliminates the distortion of differently ordered neurons and thus facilitates a successful training.
	
	Permutation invariant processing introduces one \mbox{sub-ANN} for each layer of the target model (denoted as functions $\phi_1,\dots,\phi_n$ in~\cite{ganju2018property}) in the adversarial meta classifier \advA.
	The outputs of these sub-ANNs are then combined by the function $\rho$ (another ANN) which computes the result for~\advA, i.e., the property classification of the given target model.
	%\todo{JS: Ich habe hier mal zwei Sätze rausgekürzt -- vermutlich müssen wir nicht weiter ins Detail gehen?}
	%The set-like processing is realized by calculating the sum of the parameters for a corresponding layer in the sub-NN before forwarding them to the combining NN $\rho$.
	%By training the meta classifiers solely on these sums, the ordering of neurons within a layer is ignored -- thus making the calculation permutation invariant.
	
	Experimental results in~\cite{ganju2018property} show that the performance of a PIA meta classifier \advA is significantly boosted by applying this strategy.
	
	\section{Related Work}
	\label{sec:relwork}
	This section briefly summarizes related work from the area of machine learning privacy attacks and defenses against them.
	\paragraph{PIA defense strategies}
	Effective universal defense mechanisms against PIAs have not been discovered yet~\cite{rigaki2020survey}.
	Differential privacy~\cite{dwork2006calibrating} is a promising approach against other privacy attacks like membership inference~\cite{rigaki2020survey, suri2021formalizing, suri2022subject}. However, it only slightly decreases the success rate of PIAs, since it merely limits the impact of each single input, but does not influence the presence of general properties in the training data set~\cite{ateniese2015hacking,liu2021ml, zhang2021leakage}.
	
	\citeauthor{ganju2018property}~\cite{ganju2018property} propose \emph{node multiplicative transformations} as another defense strategy.
	As long as a neural network uses ReLU or LeakyReLU as an activation function, it is possible to multiply the parameters of one layer by some constant and dividing the constants connecting it to the next layer by the same value without changing the result.
	Although they claim that this might be effective, this strategy is limited to ReLU and LeakyReLU activation functions and requires changes in the model architecture.
	
	In contrast, our \emph{\pu} approach works without any changes to the target model, i.e. the model to protect.
	It involves adversarial training, which has been successfully implemented in many domains, including privacy protection.
	
	%In this paper, however, we use a variant of machine unlearning to hide \emph{general properties} of the entire set of training data.
	
	\paragraph{Adversarial Training}
	The term ``adversarial training'' appears in two different ways in the literature:
	On the one hand, it describes the state-of-the-art defense strategy against adversarial examples~\cite{szegedy2013intriguing, athalye2018obfuscated}, an attack which manipulates data such that a target models reliably fails.
	On the other hand, the term describes the direct involvement of a modeled adversarial party during training, which is used to either improve the model's performance, e.g., in generative adversarial networks~(GANs)~\cite{goodfellow2014generative}, or to deliberately harden a model against a specific attack.
	
	In~\citeyear{nasr_machine_2018}, \citeauthor{nasr_machine_2018} have introduced the latter as \emph{adversarial regularization}, a promising strategy against black-box membership inference attacks.
	They model the training process as a min-max privacy game, during which the prediction loss of the target model is minimized, while the model is regularized to resist against (the strongest possible) membership inference attacks.
	The trade-off between utility and ``membership privacy'' is controlled by a regularization parameter.
	Experimental results suggest that this strategy is very effective, achieving robustness against membership inference at the cost of only a few percent in prediction accuracy~\cite{nasr_machine_2018}.
	
	The performance metrics of the two different forms of adversarial training are hard to compare:
	In adversarial example attacks, the adversary generally tries to deliberately weaken the utility of a target model.
	Therefore, in contrast to other (privacy) attacks like PI, there is no privacy-utility trade-off to quantify the success of attack and defense scenarios.
	Instead, it is sufficient to supply the lowest utility an adversary can achieve when applying the defense strategy, e.g., in~\cite{madry2018towards}.%\todo{JS: Brauchen wir diesen letzten Absatz?} \todo{Antwort= JW: wenn wir Platz haben, würde ich den drin lassen, der passt zu der Metrik, die wir in der Discussion erwähnen, die es zu erfinden gilt.}
	
	\paragraph{Other PIA attacks}
	\citeauthor{melis2019exploiting} explore PIAs in the context of collaborative learning:
	In this scenario, the adversary is a legitimate party in a collaborative setting, where participants jointly train a machine learning model via exchanging model updates -- without sharing their local and private data.
	The authors present an \emph{active} and a \emph{passive} method to infer a property of the training data of another participant by analyzing the shared model updates of other participants~\cite{melis2019exploiting}. 
	
	Focusing on a black-box scenario, \citeauthor{zhang2021leakage} study both single- and multi-party PIAs for tabular, text and graph data sets.
	While their attack does not need access to the parameters of a target model, several hundreds of queries to the target model are needed for the attack to be successful~\cite{zhang2021leakage}.
	
	An advanced PIA by \citeauthor{mahloujifar2022property} introduces \emph{poisoning} as a way to ease the attack in a black-box scenario.
	This requires the adversary to control parts of the training data.
	In this adversarial training data set, the label of data points with a target property \A are changed to an arbitrary label $l$.
	After training, the distribution of a target property can then be inferred by evaluating multiple queries to the target model -- loosely summarized, the more often the label $l$ is predicted, the larger the portion of samples with property \A is in the training data set~\cite{mahloujifar2022property}.
	
	\citeauthor{song2019overlearning} propose a very similar attack to property inference, which we call attribute inference:
	They assume a machine learning target model which is partly evaluated \emph{on-premise} and partly \emph{in the cloud}.
	Their attribute inference attack reveals properties of a single data instance, e.g., whether a person wears glasses on a photo during the inference phase~\cite{song2019overlearning}.
	In contrast, we focus on PIAs which reveal global properties about a whole training data set.

	\section{Property Unlearning as a Defense Strategy}
	\label{sec:pu}
	
	In this section we elaborate on our novel defense strategy against PIAs, which we call \pu.
	
	\begin{figure}[h!b]
		\centering
		\includegraphics[width=1.0\linewidth]{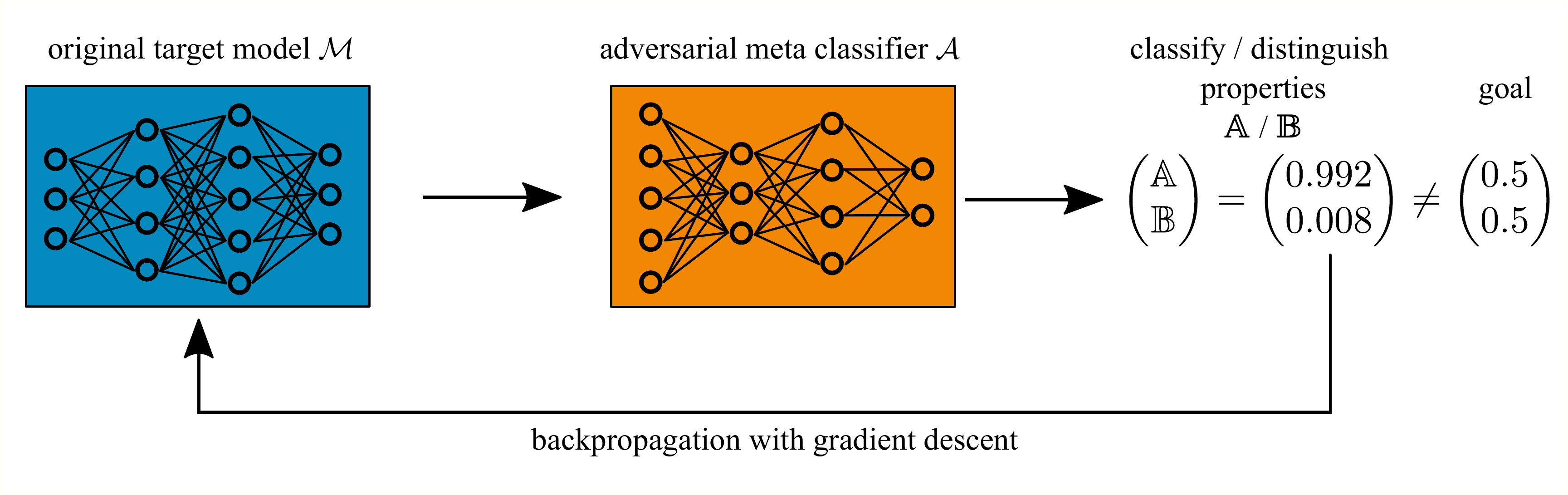}
		\caption{\puBegin as a defense strategy against PIAs. For the training of the adversarial meta classifier $\mathcal{A}$, refer to Figure~\ref{fig:pia}.}
		\label{fig:pu}
	\end{figure}
	
	As a prerequisite, an adversarial classifier \advA needs to be constructed.
	This is achieved as described in Section~\ref{sec:pia}:
	constructing one auxiliary data set~$DS$ for each property \A and \B, and training a set of \emph{shadow models} for each property with the corresponding data sets \DSA and \DSB.
	Note that when creating an adversary as a preparation for protecting one's own model, the auxiliary data sets $DS_\mathbb{A}$ and $DS_\mathbb{B}$ can trivially be subsets of the original training data of the target model, since the model owner has access to the full training data set.
	This yields in very strong adversarial accuracy as opposed to an outside adversary who might need to approximate or extract this training data first.
	The same holds for white-box access to the model, which is straightforward for the owner of a model.
	Hence, the training of a reasonably good adversarial meta classifier \advA ($>99\%$ test accuracy) as a first step of \pu is easily achievable for the model owner (see Section~\ref{subsec:data sets}).
	As a second prerequisite, the target model \targetM, which the owner wants to protect, also needs to be fully trained with the original training data set -- having either property \A or \B.
	
	To \emph{unlearn} the property from \targetM, we use backpropagation.
	Just like during the regular training process, the parameters of the target model \targetM\ are modified by calculating and applying gradients.
	But different from the original training, \pu does not optimize \targetM towards better classification accuracy.
	Instead, the goal is to disable the adversary \advA from extracting the property \A or \B from \targetM.
	
	In practice, the output of the adversarial meta-classifier \advA is a vector of length 2 (or: number of properties~$k$) which sums up to 1.
	Each value of the vector corresponds to the predicted probability of a property. 
	As an example, the output $[0.923, 0.077]$ means that the adversary \advA is 92.3\% confident that \targetM has property \A, and only 7.7\% to have property \B.
	Thus, \pu aims to disable the adversary from making a meaningful statement about \targetM, i.e., an adversary output of $[0.5, 0.5]$ is pursued -- or more generally $[\frac{1}{k}, \dots, \frac{1}{k}]$ for $k$ properties.
	
	Algorithm~\ref{alg:property-unlearning} shows pseudocode for the \pu algorithm.
	As expected, the termination condition for the while-loop in line~\ref{algline:termination} addresses the ability of the adversary \advA:
	As long as \advA is significantly more confident for one of the properties, the algorithm needs to continue. 
	
	After calculating the gradients~$g$ automatically via TensorFlow's~\cite{tensorflow} backtracking algorithm in line~\ref{algline:calc-grad}, the actual unlearning happens in line~\ref{algline:do_unlearning}.
	Here, the gradients are applied on the parameters of model~\targetM, nudging them to be less property-revealing.	
	As described in Section~\ref{sec:ML-basics}, the learning rate controls how large the impact of the gradients should be in a single step.
	If the parameters have been changed \emph{too much}, the current $\mathcal{M}'$ gets discarded and the gradients are reapplied with a smaller learning rate (see line~\ref{algline:decr_lr}; also visualized in Figure~\ref{fig:adv-utility}). 
		
	The effect of \pu in between rounds of the algorithm is measured by the \emph{adversarial utility}, see lines~\ref{algline:advut-start}--\ref{algline:advut-end}.
	We calculate the adversarial utility by analyzing the adversary output $Y$.
	Recall that $Y$ is a vector with $k$ entries, with each entry $Y_i$ representing the adversarially estimated probability that the underlying training data set of the target model $\mathcal{M}$ has property $i$.
	The adversarial utility is defined by the largest absolute difference of an entry $Y_i$ to $\frac{1}{k}$ (see line~\ref{algline:advut}).
	Remember that the goal of \pu is to nudge the parameters of $\mathcal{M}$ such that the output of the adversary is close to $\frac{1}{k}$ for all $k$ entries in the output vector $Y$.
	The condition in line~\ref{algline:if-adv-ut-dec} therefore checks whether the last parameter update from $\mathcal{M}$ to $\mathcal{M}'$ was useful, i.e., whether the adversarial utility has decreased.
	Only if this is the case, the algorithm gets closer to the \pu goal.
	Otherwise, the last update in $\mathcal{M}'$ is discarded and the next attempt is launched with a lower learning rate.
	
	\begin{algorithm}
		\caption{\puBegin for a target model \targetM, using property inference adversary \advA, initial learning rate~$lr$, and set of properties $P = \{\mathbb{A}, \mathbb{B}, ...\}$}
		\label{alg:property-unlearning}
		\begin{algorithmic}[1]
			
			\Procedure{PropertyUnlearning}{$\mathcal{M}$, \advA, $lr$, $P$}
			\State $k \gets |P|$ \Comment{number of properties (default 2)}
			\State $Y \gets \mathcal{A}(\mathcal{M})$ \Comment{original adv. output with $|Y| = k$}
			\State let $i \in [k]$
			\While{$\exists i: Y_i \gg \frac{1}{k}$ or $Y_i \ll \frac{1}{k}$} \label{algline:termination}
			%\While{$\exists i: Y_i = 1$} \Comment{Perfect adv. confidence}
			%\label{algline:while-perf-start}
			%\State $\mathcal{M} \gets$ randomly flip 0.2\% of $\mathcal{M}$'s weights
			%\State $Y \gets \mathcal{A}(\mathcal{M})$ \Comment{Update adversarial output}
			%\EndWhile
			%\label{algline:while-perf-end}
			\State $g \gets$ gradients for $\mathcal{M}$ s.t. $\forall i: Y_i \to \frac{1}{k}$
			\label{algline:calc-grad}
			\State $\mathcal{M}' \gets$ apply gradients $g$ on $\mathcal{M}$ with $lr$ \label{algline:do_unlearning}
			\State $Y' \gets \mathcal{A}(M')$ \Comment{update adversarial output}
			\If{\Call{AdvUtility}{$Y'$} $<$ \Call{AdvUtility}{$Y$}}\label{algline:if-adv-ut-dec}
			\State $\mathcal{M}, Y \gets \mathcal{M}', Y'$
			\Else
			\State $lr \gets$ $lr/2$ \label{algline:decr_lr} \Comment{retry with decreased $lr$}
			\EndIf
			\EndWhile
			\State \Return $\mathcal{M}$
			\EndProcedure
			
			\Function{AdvUtility}{adversarial output vector $Y$}
			\label{algline:advut-start}
			\State $k \gets |Y|$ \Comment{number of properties (default 2)}
			\State \Return $\underset{i \in [k]}{\text{max}}(|Y_i-\frac{1}{k}|)$ \Comment{biggest difference to $\frac{1}{k}$}
			\label{algline:advut}
			\EndFunction
			\label{algline:advut-end}
		\end{algorithmic}
	\end{algorithm}
	
	\section{\puBegin experiments}
	\label{sec:pu-exp}
	%\subsection{Evaluation setup - ggf. Überschrift einfach weglassen}
	%\label{sec:exp-setup}
	To test the \pu approach in practice, we have conducted extensive experiments with different data sets.
	First, we explain our implementation of the adversarial meta classifier~\advA, which serves as the attacker for our defense approach.
	Second, the data sets to unlearn one property are introduced in detail.
	We then continue by describing our experiments and their results.
	
	We have also conducted a multi-property unlearning experiment (aimed at unlearning multiple properties from a target model) which led to similar results and is therefore omitted in this paper -- see Appendix \ref{appendix-multi-prop}.
	
	\subsection{Adversarial property inference classifier}
	\label{sec:pia-adv}
	As described in Section~\ref{sec:pia}, we use the permutation invariant attack approach by \citeauthor{ganju2018property}~\cite{ganju2018property}.
	Per data set, we train one adversarial meta classifier network \advA, which is able to extract the respective properties~\A and~\B from a given target model.
	
	Depending on the number of neurons in a layer of the target model, the sub-NNs $\phi$ consist of 1--3 layers of dense-neurons, containing 4--128 neurons each.
	In the adversarial meta classifier \advA, the number of layers and number of neurons within the layers are proportionate to the input size, i.e., the number of neurons in the layer of the target model.
	These numbers are evaluated experimentally, such that the meta classifiers perform well, but do not offer more capacity than needed (which would encourage overfitting).
	
	The sub-network $\rho$ of the adversarial meta classifier consists of 2--3 dense-layers with 2--16 dense-neurons each.
	In our experiments the output layer always contains two neurons, one for each property~\A and~\B.
	
	For each of the three data sets in the next section, we apply the following steps to prepare for \pu:
	\begin{itemize}
		\item Develop a target model \targetM which performs well on the given classification task.
		\item Extract two auxiliary data sets $DS_\mathbb{A}$ and $DS_\mathbb{B}$ for each property~\A and~\B.
		\item Use each of the auxiliary data sets $DS_\mathbb{A}$ and $DS_\mathbb{B}$ as training data for \num{2000} shadow models. Recall that the shadow models have the same architecture as the target model \targetM.
		\item Develop and train an adversarial meta classifier \advA on the parameters of the shadow models, performing well in extracting the given properties.
	\end{itemize}
	
	We then employ this adversarial model \advA in our \pu algorithm, as described in Section~\ref{sec:pu}.
	
	\begin{table*}[t]
		\centering
		\caption{The data sets used for the experiments. avg=average, distrib.=distribution, PIA=Property Inference Attack, Acc.=Accuracy, \mbox{PU=Property Unlearning.}}
		\begin{tabular}{ccccccc}
			\toprule
			\multirow{3}{*}{\textbf{Experiment}} &
			%\multirow{2}{*}{\textbf{Data Set}} &
			\multirow{3}{*}{\textbf{Size}} &
			\multirow{3}{1.5cm}{\centering \textbf{Task}} &
			\multirow{3}{*}{\textbf{Target Property}} &
			\multirow{3}{*}{\textbf{$|DS_{*}|$}} &
			\multirow{3}{1.5cm}{\centering \textbf{Initial PIA Accuracy}} &
			\multirow{3}{2.5cm}{\centering \textbf{Avg. Task Acc. Loss due to PU}} \\
			\addlinespace[18pt]
			%\midrule
			\midrule
			\expMnist %& MNIST
			& 70K & digits & noise & 12K  &  100\% &  0.6\%P \\
			%\midrule
			\midrule
			\expCensus %& Census
			& 48K & income & gender distrib. & 15K &  99.3\% & 0.2\%P  \\
			\midrule
			%\midrule
			\expUtk %& UTKFace
			& 23K & gender & race distrib. & 10K & 99.8\% & 0.8\%P \\
%			\midrule
%			\expMultiAR, \expMultiRA %& UTKFace (multi)
%			& 23K & gender & race / age distrib. & 10K & 91--98\% & 0.9--1.7\%P \\
			\bottomrule
		\end{tabular}
		\label{table:data sets}
	\end{table*}
	
	\subsection{Data sets and network architectures}
	\label{subsec:data sets}
	We use three different data sets to evaluate our approach, as summarized in Table~\ref{table:data sets}.
	For each data set and auxiliary data set $DS_*$, we train \num{2000} shadow models and \num{2000} target models.
	While the shadow models are used to train the adversaries $\mathcal{A}$, the target models $\mathcal{M}$ are the subjects of our experiments, i.e., we apply \pu on these target models and measure the resulting privacy-utility trade-off.
	The shadow models and target models share the same architecture per data set.
	
	\textbf{MNIST}~\cite{lecun1998gradient} is a popular database of handwritten digit images.
	It consists of \num{70000} labeled images of the digits 0--9.
	Similar to \citeauthor{ganju2018property}~\cite{ganju2018property}, we distort all images with Gaussian noise (parameterized with $\text{mean}=35$, $\text{sd}=10$) in a copy of the database.
	For the MNIST property inference attack, we choose the property of having original pictures without noise ($\mathbb{A}_\textit{MNIST}$) and pictures with noise ($\mathbb{B}_\textit{MNIST}$).
	Our models for the MNIST classification task are ANNs with a preprocessing-layer to flatten the images, followed by a 128-neurons dense-layer and a 10-neuron dense-layer for the output.
	We train with a batch size of \num{128} for six epochs.
	On average, our shadow models trained for MNIST digit classification yield a test accuracy of 94.5\% if trained with the original data set $DS_{\mathbb{A}_\textit{MNIST}}$ and 88.3\% when trained with the noisy data set $DS_{\mathbb{B}_\textit{MNIST}}$.
	Note that for performance reasons (faster training of the shadow models) and a more realistic scenario, each of the two data sets $DS_*$ only have a size of \num{12000}.
	
	The adversarial meta classifier \advA$_\textit{MNIST}$\ that was trained against these MNIST models is very efficient, reaching 100\% test accuracy.
	This means that it correctly distinguishes between models trained with $DS_{\mathbb{A}_\textit{MNIST}}$ and models trained with $DS_{\mathbb{B}_\textit{MNIST}}$ at all times.
	
	\textbf{Census}~\cite{Dua2019} (\emph{Census Income Data Set}, also called ``Adult Data Set'') is a tabular data set containing more than \num{48000} instances for income prediction.
	With 14 attributes, the task is to determine for each person (resp. data instance) whether they earn over \num{50000}\$ a year.
	The property inference attack aims at extracting the ratio of male to female persons in the database, which is originally 2:1.
	We select \num{7500} female persons and just as many male instances from the original Census data set to create a data set for property~$\mathbb{A}_\textit{Census}$, resulting in a male:female ratio of 1:1.
	Similarly, the data set for property~$\mathbb{B}_\textit{Census}$ consists of \num{10000} male instances and \num{5000} female instances, with the original ratio of 2:1.
	The architecture of the Census models consists of one 20-neurons dense-layer and a 2-neurons output dense-layer.
	We achieve an accuracy of 84.7\% on average, both when training with $DS_{\mathbb{A}_\textit{Census}}$ and with $DS_{\mathbb{B}_\textit{Census}}$.
	The PIA model $\mathcal{A}_\textit{Census}$ distinguishes the described Census models with a test accuracy of 99.3\%.

	\textbf{UTKFace}~\cite{zhang2017age} contains over \num{23000} facial images.
	Each image is labeled with the three attributes age, gender and ethnicity (with five possible values White, Black, Asian, Indian, and Others).
	We choose gender recognition as the task for the target models~\targetM.
	Therefore, we create a data set consisting only of images with ethnicity \emph{White} from the original data set for property~$\mathbb{A}_\textit{UTK}$.
	Accordingly, the data set for property~$\mathbb{B}_\textit{UTK}$ is comprised of images labeled with \emph{Black}, \emph{Asian}, \emph{Indian}, and \emph{Others}.
	Both auxiliary data sets $DS_{\mathbb{A}_\textit{UTK}}$ and $DS_{\mathbb{A}_\textit{UTK}}$ have a size of \num{10000}~images.
	
	For the UTKFace gender recognition task, we use a convolutional neural network~(CNN) architecture with three sequential combinations of convolutional, batch normalization, max-pooling and dropout layers, leading to one dense-layer with 2~neurons generating the output.
	After training the models for 33~epochs, we achieve a test accuracy of 88.0--88.3\% on average for gender classification, depending on the property of the respective training data set.
	Our property inference adversary $\mathcal{A}_\textit{UTK}$ against the target models $\mathcal{M}_\textit{UTK}$ has 99.8\% test accuracy in distinguishing the properties of $\mathcal{M}_\textit{UTK}$.
	
	% In 1--2 Sätzen: Auch multi property experiments beschreiben
	
	\subsection{Experiment 1: Single \pu}
	\label{sec:exp1-results}
	In this section we experimentally evaluate the performance of \pu to defend against a specific PIA adversary.
	For each of the data sets described above, we have trained \num{2000} test models in the same way we have created the shadow models.
	We refer to these test models as target models.
	In all of the experiments, \pu is deployed successfully and achieves a good privacy-utility trade-off.
	
	The figures in this section contain boxplot-graphs.
	Each boxplot consists of a box, which vertically spans the range between the first quartile $Q_1$ and the third quartile $Q_3$, i.e., the range between the median of the lower half and the upper half of the data set.
	This means that half of the data points are covered by the box.
	The ``whiskers'' are vertical lines above and below the box.
	The lower whisker spans the range between the first quartile and 1.5 times the interquartile range (IQR), defined by the difference between the third and first quartile.
	Therefore, the lower whisker spans data points between $Q_1$ and $Q_1-1.5*\text{IQR}$ with IQR = $Q_3-Q_1$.
	Analogously, the upper whisker spans data points between $Q_3$ and $Q_3+1.5*\text{IQR}$.
	All points which are neither covered by the box nor by its whiskers are outliers and represented by asterisks.
	The horizontal line in a box marks the median and the diamond marker indicates the average value across the displayed data set.

\begin{figure*}[t]
		\centering
		\subfloat[\expMnist before and after \pu\label{fig:mnist-eff}]{
			\includegraphics[height=1.4\figheight]{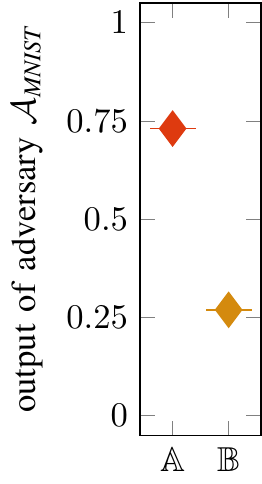}
			\includegraphics[height=1.4\figheight]{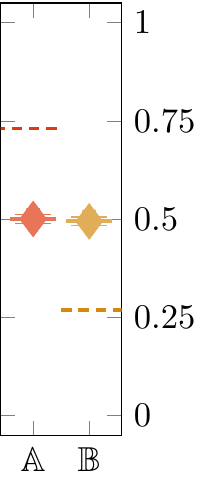}}
		\hfill
		\subfloat[\expCensus before and after \pu\label{fig:census-eff}]{
			\includegraphics[height=1.4\figheight]{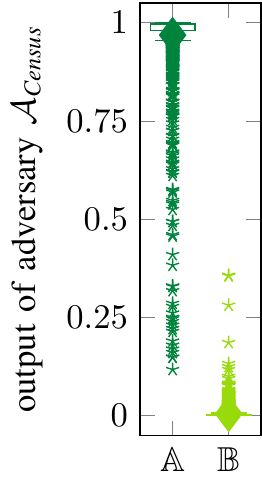}
			\includegraphics[height=1.4\figheight]{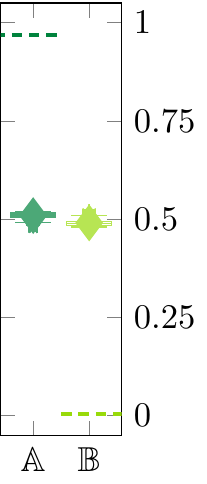}}
		\hfill
		\subfloat[\expUtk before and after \pu\label{fig:utk-eff}]{
			\includegraphics[height=1.4\figheight]{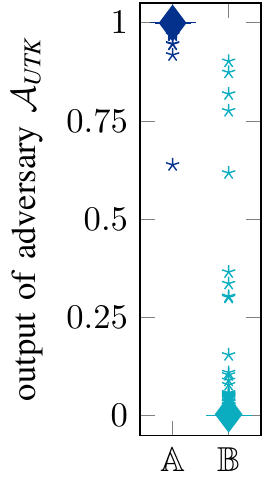}
			\includegraphics[height=1.4\figheight]{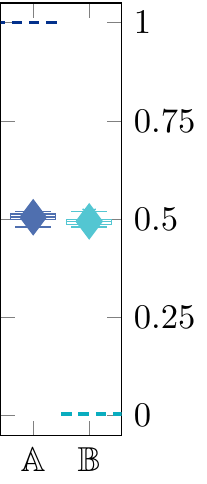}}
		\caption{Efficiency of \pu in the three experiments \expMnist, \expCensus and \expUtk.
			Each figure depicts the certainty of the adversary \advA\ in classifying \A and \B. For each sub-figure, while the left side shows the original state before, the right side shows the outputs after \pu.
			In each experiment, both properties $\mathbb{A}$ and $\mathbb{B}$ are unlearned successfully from the target models $\mathcal{M}$. The dashed lines represent the average accuracy values \emph{before} \pu was applied. Each boxplot visualizes the values of 2000 target models.
		}
	\end{figure*}

\begin{figure*}[h!t]
		\centering
		\subfloat[\expMnist before and after \pu\label{fig:mnist-acc}]{
			\includegraphics[height=1.4\figheight]{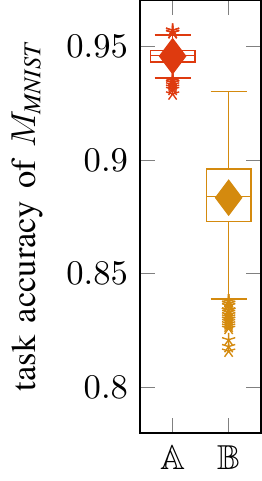}
			\includegraphics[height=1.4\figheight]{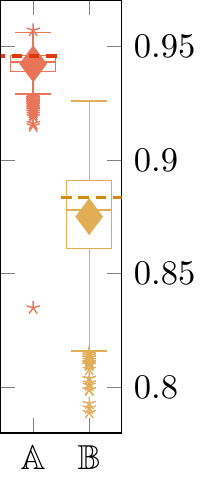}}
		\hfill
		\subfloat[\expCensus before and after \pu\label{fig:census-acc}]{
			\includegraphics[height=1.4\figheight]{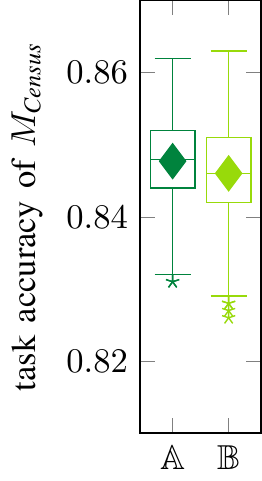}
			\includegraphics[height=1.4\figheight]{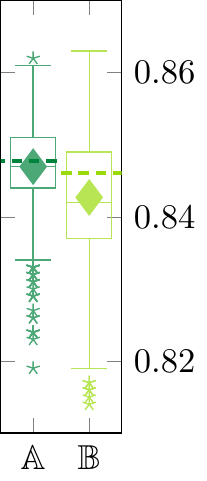}}
		\hfill
		\subfloat[\expUtk before and after \pu\label{fig:utk-acc}]{
			\includegraphics[height=1.4\figheight]{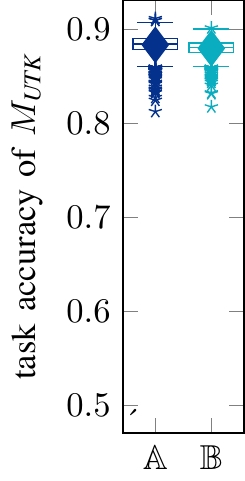}
			\includegraphics[height=1.4\figheight]{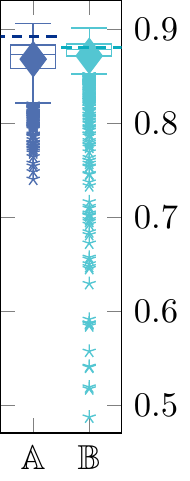}}
		\caption{The results of the experiments \expMnist, \expCensus and \expUtk regarding the accuracy loss of the target models $\mathcal{M}$ due to \pu. The dashed lines represent the average accuracy values \emph{before} \pu was applied. Each boxplot visualizes the values of 2000 target models.}
	\end{figure*}

	\paragraph{MNIST} For the MNIST experiment \expMnist, the adversary classifies the properties \A and \B with high certainty in all instances before unlearning, see Figure~\ref{fig:mnist-eff}.
	After unlearning, the adversary cannot infer the property of any of the MNIST target models $\mathcal{M}_\textit{MNIST}$ -- as intended.
	Meanwhile, the accuracy of the target models $\mathcal{M}_\textit{MNIST}$ decreased slightly from an average of 94.6\% by 0.4\%P to 94.2\% for models with property \A, respectively from 88.3\% by 0.8\%P to 87.5\% for models with property \B\ (see Figure~\ref{fig:mnist-acc}).
	Recall that property \B\ was introduced by applying noise to the training data, hence the affected models perform worse in general.
	
	\paragraph{Census} Similar to MNIST, \pu was successfully applied in the \expCensus experiment to harden the target models $\mathcal{M}_\textit{Census}$ against the PI adversary $\mathcal{A}_\textit{Census}$, see Figure~\ref{fig:census-eff}.
	Note that the performance of $\mathcal{A}_\textit{Census}$ is not ideal for property \A, classifying some of the instances incorrectly.
	However, 99.3\% of the \num{2000} instances were classified correctly by the adversary \emph{before} \pu.
	As desired, the output of $\mathcal{A}_\textit{Census}$ is centered around 0.5 for both properties after \pu.
	%\todo{can be provide exact valus here or maybe in a separate table somewhere?}
	The magnitude of the target models' accuracy loss is small, with an average drop of 0.1\%P for property \A~(84.8\% to 84.7\%) and 0.3\%P~(84.6\% to 84.3\%) for property \B, see Figure~\ref{fig:census-acc}.
	
	\paragraph{UTKFace} Moreover, in the \expUtk experiment, \pu could be successfully applied to all models~(see Figure~\ref{fig:utk-eff}) to harden the target models against PIAs.
	On average, the accuracy of the target models dropped by 1.3\%P from 88.2\% to 86.9\% for models trained with the data set $DS_\mathbb{A}$ and by 0.1\%P from 87.9\% to 87.8\% for target models trained with $DS_\mathbb{B}$, see Figure~\ref{fig:utk-acc}.
	This yields an average accuracy drop of 0.8\%P across the target models for both properties (from 88.1\% to 87.3\%).
	%\todo{maybe write a sentence on the adversary that goes to 0.5?}
	
	\subsection{Experiment 2: Iterative \pu}
	\label{sec:exp2-results}
\begin{figure}%[h!b]
		\centering
		\includegraphics[width=1.0\linewidth]{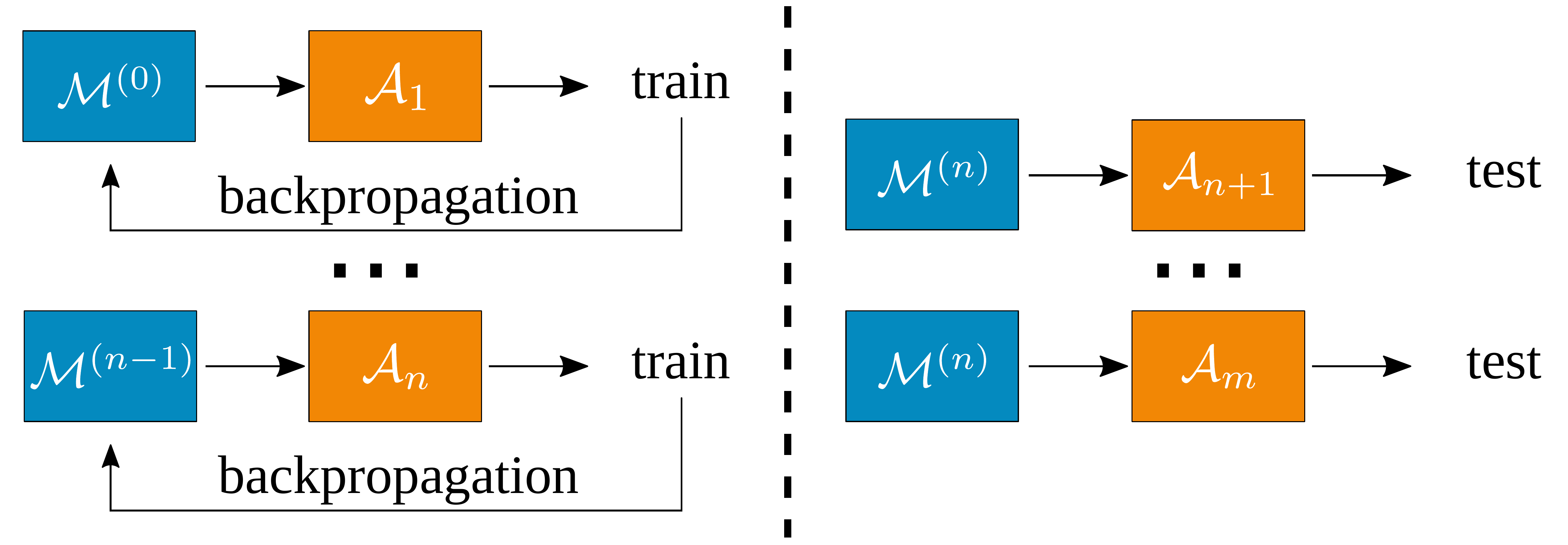}
		\caption{In reference to Figure~\ref{fig:pu}, iterative \pu works by performing single \pu for $n$ different adversarial meta classifiers $\mathcal{A}$ iteratively on the original target model $\mathcal{M}^{(0)}$. Afterwards, the resulting target model $\mathcal{M}^{(n)}$ can be evaluated by testing the unlearned target model my additional $m$ adversarial meta classifiers.}
		\label{fig:iterpu}
	\end{figure}

	% keine Be/Auswertung des ersten Experiments, nur beschreiben, wie das zweite Experiment aufgebaut ist und wie die Ergebnisse zu verstehen sind
	In the previous section, the results of Experiment 1 have shown that \pu can harden a target model~\targetM against a single PI adversary, i.e., a specific adversarial meta classifier $\mathcal{A}$ (see Figure \ref{fig:pu}).
	The setup of Experiment 2 aims to improve that by generalizing the unlearning. Therefore, the same target model $\mathcal{M}$ is unlearned iteratively against a range of different adversaries $\mathcal{A}$ (see Figure \ref{fig:iterpu}). 
	The results of our experiments are based on 200 target models.
	We unlearned each initial target model $\mathcal{M}^{(0)}$ iteratively for $n$ different adversarial meta classifiers $\mathcal{A}_i$, where $n = 15$. 
	After that, the resulting iteratively unlearned target model $\mathcal{M}^{(n)}$ is tested by another distinct adversarial meta classifier. 
	To increase the significance of our results, we choose to test the resulting target model $\mathcal{M}^{(n)}$ with $m = 5$ additional distinct adversarial meta classifiers. 
	Furthermore, we apply a 4-fold cross validation technique to this constellation of in total 20 distinct adversarial classifiers.  
	Finally, the results are plotted in boxplots similar to Experiment 1:
	Here, each boxplot is visualizing $200 \text{~(target models)} * 4 \text{~(folds)} * 5 \text{~(adversary outputs in a fold)} = 4000$ data points.
	
	The shadow models which were used to train the 20 adversaries $\mathcal{A}$ have been grouped in a way to make sure that the 5 testing adversaries' training set is disjunct from the training set of the 15 adversaries used for unlearning.
	The order of the 15 adversaries for unlearning has been chosen randomly for each of the 200 target models $\mathcal{M}$.

\begin{figure}%[h!b]
		\centering
		\includegraphics[height=0.7\figheight]{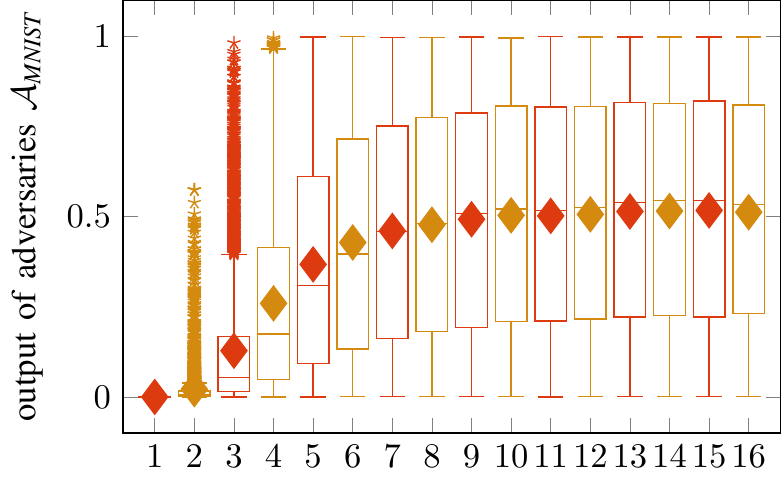}
		\includegraphics[height=0.7\figheight, trim=19 0 0 0, clip]{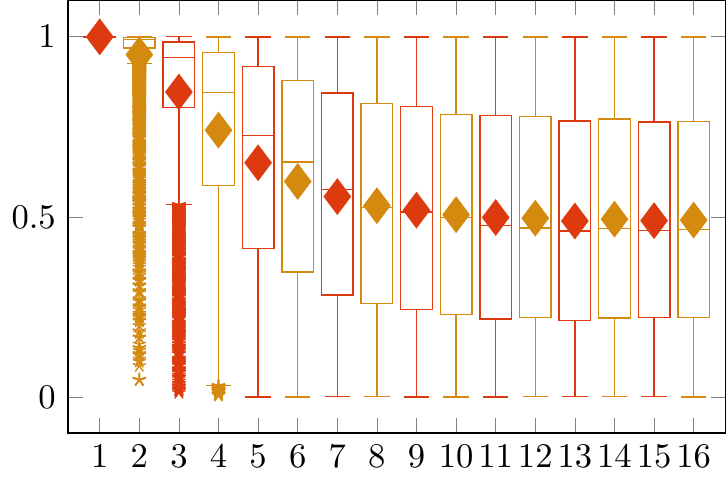}
		\caption{Results of iterative unlearning experiment for property \A (left) and \B (right). For each of the 200 target models $\mathcal{M}$, the predictions of all 5 testing adversaries are plotted along the y-axis before unlearning (first column) and after each unlearning iteration (other 15 columns).}
		\label{fig:cross1}
	\end{figure}

\begin{figure}
		\centering
		\subfloat[\label{subfig:cross2ag1}]{
		\includegraphics[height=1.1\figheight]{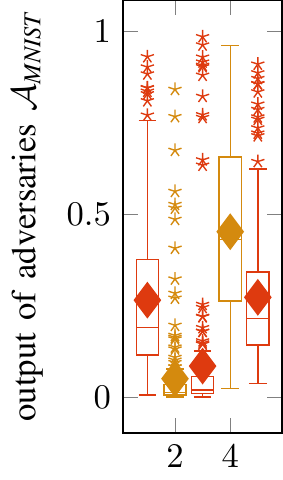}
			}
		\hfill
		\subfloat[\label{subfig:cross2ag2}]{
		\includegraphics[height=1.1\figheight, trim=19 0 0 0,clip]{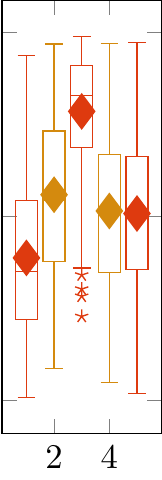}
			}
		\hfill
		\subfloat[\label{subfig:cross2ag3}]{
		\includegraphics[height=1.1\figheight, trim=19 0 0 0,clip]{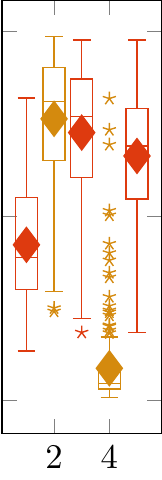}
			}
		\hfill
		\subfloat[\label{subfig:cross2ag4}]{
		\includegraphics[height=1.1\figheight, trim=19 0 0 0,clip]{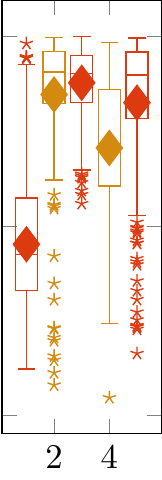}
		}
		\caption{Individual adversary outputs after all 15 unlearning iterations for property \A target models. Recall that before unlearning, all adversaries have correctly inferred property \A by outputting $y=0$.}
		\label{fig:cross2a}
	\end{figure}

\begin{figure}
		\centering
		\subfloat[\label{subfig:cross2bg1}]{
		\includegraphics[height=1.1\figheight]{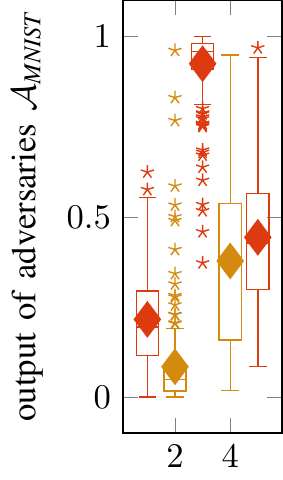}
			}
		\hfill
		\subfloat[\label{subfig:cross2bg2}]{
		\includegraphics[height=1.1\figheight, trim=19 0 0 0,clip]{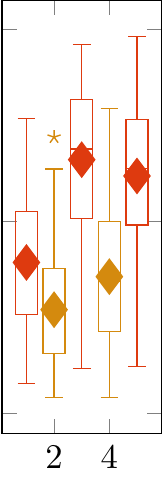}
			}
		\hfill
		\subfloat[\label{subfig:cross2bg3}]{
		\includegraphics[height=1.1\figheight, trim=19 0 0 0,clip]{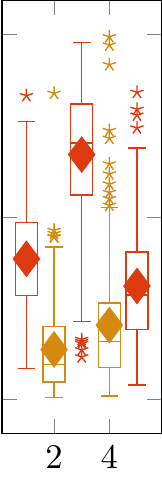}
			}
		\hfill
		\subfloat[\label{subfig:cross2bg4}]{
		\includegraphics[height=1.1\figheight, trim=19 0 0 0,clip]{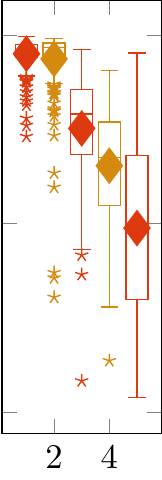}
		}
		\caption{Individual adversary outputs after all 15 unlearning iterations for property \B target models. Recall that before unlearning, all adversaries have correctly inferred property \B by outputting $y=1$.}
		\label{fig:cross2b}
	\end{figure}

	The overall results of this experiment on the MNIST data set are presented in Figure \ref{fig:cross1}. They show the iterative unlearning process for property~$\mathbb{A}$ (left-hand side) and property~$\mathbb{B}$ (right-hand side). Each column on the x-axis represents an iteration step of the iterative unlearning procedure. On the y-axis, the prediction of the adversary regarding the corresponding property is plotted, which is ideal for property~$\mathbb{A}$ and $\mathbb{B}$ to be 0 and 1, respectively. $y=0.5$ is the goal of the property unlearning defense strategy, such that the attacker is not able to distinguish a certain property for a given data set. 
	Clearly, the second column shows that after applying \pu once, a distinct adversary (i.e., not the adversary which was involved in the unlearning process) is still able to sufficiently infer the correct property for most target models~\targetM.
	The plots show that after about ten iterations of property unlearning, the average output of the 5 testing adversaries converges towards an average of prediction probability 0.5 (for both properties \A and \B).
	While this could be misinterpreted as ultimately reaching the goal of property unlearning, we introduce Figures \ref{fig:cross2a} and \ref{fig:cross2b} that paint a more fine-grained picture of the last column of Figure \ref{fig:cross1}. 
	Here, each of the four plots contain five independent boxplots corresponding to the five distinct test adversaries in one fold of the cross validation process.
	Each boxplot presents the prediction results of one adversary for the 200 independently unlearned target models $\mathcal{M}^{(15)}_i$ of the experiment.
	
	While the plots of Figure \ref{fig:cross1} suggest that the adversaries' outputs are evenly spread across the interval $[0,1]$ with both an average and median close to 0.5, Figures \ref{fig:cross2a} and \ref{fig:cross2b} show that this is only true for the indistinct plot of all 4 experiments with 5 testing adversaries each.
	We want to point out three key observations:
	\begin{enumerate}
		\item The majority of adversaries does \emph{not} have a median or average output near 0.5 after 15 unlearning iterations for the 200 target models $\mathcal{M}$.
		\item For some adversaries $\mathcal{A}$, the target models have been ``over-unlearned'' during the 15 iterations with their output clearly nudged into opposite of their original output, e.g., the third adversary in Figure~\ref{subfig:cross2ag4}, or the second adversary in Figure~\ref{subfig:cross2bg3}. 
		\item Most importantly, other adversaries are still correctly inferring the property for most or even all 200 target models with high confidence after the 15 unlearning iterations, e.g., the second adversary in Figure~\ref{subfig:cross2ag1}, or the first two adversaries in Figure~\ref{subfig:cross2bg4}.
	\end{enumerate}
	
	\begin{figure}%[t]
		\centering
		\includegraphics[height=\figheight]{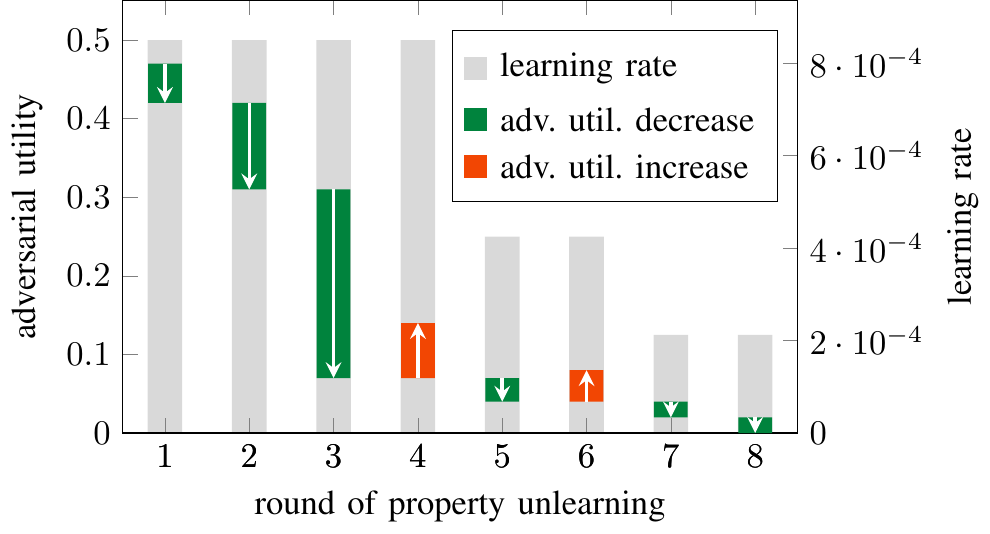}
		\caption{A visualized example of the decreasing adversarial utility during property unlearning with \emph{one} adversary for a single target model $\mathcal{M}$. In each round, the adversarial utility of $\mathcal{M}$ either decreases further towards the goal of~¸0 (green bar), or the unlearning round is repeated with a smaller learning rate (after a red bar). The final result of round 8 is a completely \emph{unlearned} target model $\mathcal{M}$ with an adversarial utility close to 0, see Algorithm~\ref{alg:property-unlearning}.}
		\label{fig:adv-utility}
	\end{figure}
	
	\subsection{Time costs}
	As a prerequisite to \pu, the training of the adversary \advA\ can be a time consuming process, especially the training of the shadow models.
	Since the shadow models are trained on smaller data sets (see Table~\ref{table:data sets}), training takes slightly less time.
	With a 14-core Intel~Xeon E5-2660~CPU and a clock speed of 2.0GHz, training \emph{one} shadow model takes about 8~seconds for the fully connected neural networks we used for MNIST as well as Census, and about 5~minutes in case of the convolutional neural networks we used for UTKFace.
	
	\puBegin itself takes less time than training a shadow model.
	For MNIST and Census, the unlearning process takes about 3 to 15 seconds on average, while it takes 20 to 30 seconds on average for UTKFace target models.
	The amount of time that is consumed for \pu highly depends on the number of \emph{rounds} the algorithm requires, i.e., the amount of runs of the while-loop starting in line~\ref{algline:termination} of Algorithm~\ref{alg:property-unlearning}.
	In turn, the number of rounds depends on many factors, e.g., the magnitude of the initial adversarial utility, the number of times the learning rate has to be corrected. Typically, we observed that \pu requires between 2 and 20 rounds to unlearn a property.
	A visualization of an exemplary run is given in Figure~\ref{fig:adv-utility}.
	
	\subsection{Experiment comparison and discussion}
	
	Recall our goal for \pu: 
	We wanted to harden target models in a generic fashion, such that arbitrary PI adversaries are not able to infer pre-specified properties after applying \pu.
	
	Experiment~1 (single property unlearning) shows that \pu is very reliable to harden target models against specific adversaries.
	However, Experiment~2 (iterative \pu) indicates that single \pu fails to generalize, i.e., protect against all PI adversaries of the same class.
	This is shown in Experiment 2 by putting each target model through 15 iterations of \pu with one distinct adversary per iteration.
	After this, some adversaries are still able to infer the original properties of all target models (see third key observation in Section~\ref{sec:exp2-results}).
	This means that in the worst case, i.e., for the strongest adversaries, 15 iterations of \pu do not suffice -- while for other (potentially weaker) adversaries, 15 or even less iterations are enough to harden the models against them.
	In conclusion, \pu does not meet our goal of being a generic defense strategy, i.e., protecting not against specific adversaries but a whole class of adversaries.	
	
	\section{Experiment 3: LIME}
	\label{sec:lime}
	To explore the reasons behind this limitation of \pu, we use the explainable~AI tool by \citeauthor{ribeiro2016should}~\cite{ribeiro2016should}:
	LIME~(\emph{Local Interpretable Model-agnostic Explanations}) allows to analyze decisions of a black-box classifier by permuting the values of its input features. 
	By observing their impact on the classifier's output, LIME generates a comprehensible ranking of the input features.
	
	Recall that in the previous experiment (Section~\ref{sec:exp2-results}), we have seen that adapting the weights of a target model such that an adversarial meta-classifier \advAone cannot launch a successful PIA does not defend against another adversarial meta-classifier \advAtwo trained for the same attack.
	Therefore, we use LIME to see whether different adversarial meta-classifiers \advAone and \advAtwo rely on the same weights of a target model~\targetM to infer a property \A or \B.
	
	For comprehensible results, we use LIME images.
	We convert the trained parameters of an MNIST target model \targetM into a single-dimensional vector with a length of \num{101770}, so LIME can interpret them as an image.
	For segmentation, we use a dummy segmentation algorithm which treats each weight of \targetM (resp. pixel) as a separate segment of the 'image'. 
	This is necessary because unlike in an image, neighboring 'pixels' of \targetM's weights do not necessarily have semantic meaning.
	For reproducible and comparable results, we have initialized each LIME instance with the same random seed.
	
	\subsection*{LIME results}
\begin{figure}%[h!b]
		\centering
		\includegraphics[width=0.45\linewidth]{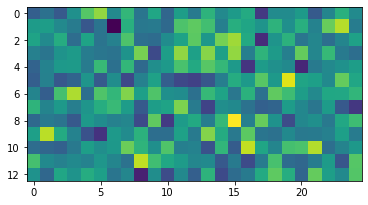}
		\includegraphics[width=0.45\linewidth]{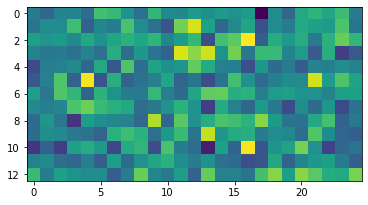}
		\caption{Two (partial) heat maps produced by LIME~\cite{ribeiro2016should} to explain the property inference of two different meta-classifiers \advAone (left) and \advAtwo (right) for the same MNIST target model~\targetM. Since \targetM has more than 100K parameters, only the first 784 are depicted here. Dark blue pixels represent parameters which have a high impact on the decision of \advA, yellow pixels represent parameters with a low impact.}
		\label{fig:lime}
	\end{figure}

	We have instantiated LIME with two property inference meta-classifiers \advAone and \advAtwo to explain their output for the same MNIST target model instance \targetM.
	The output of LIME is a heat map representing the weights and biases of \targetM, see Figure~\ref{fig:lime}.
	For practical reasons, we have only visualized the first 784 pixels of the heat map and transformed them to a two-dimensional space.
	Although \advAone and \advAtwo are trained in the same way and with the same shadow models (see Section~\ref{sec:pu-exp}), the two heat maps for classifying the property of the same target model~\targetM in Figure~\ref{fig:lime} are clearly different:
	While some of the weights have similar importance (i.e. the heat map pixels have a similar color), many weights -- especially most important (dark blue) and least significant (yellow) weights -- have very different importance for the two adversarial meta classifiers \advAone and~\advAtwo.
	
	%\todo[inline]{andere Datensätze ausprobieren, UTKFace und  Census}
	
	\section{Experiment 4: t-SNE visualization}
	\label{sec:t-sne}
	
	To understand why meta-classifiers can rely on different parts of target model parameters to infer a statistical training data property, we set out to analyze the differences in the parameters induced by such properties on an abstract level.
	
	The technique t-Distributed Stochastic Neighbor Embedding (t-SNE) \cite{van2008visualizing} is a form of dimensionality reduction which is useful for the clustering and visualization of high-dimensional data sets.
	In particular, the algorithm needs no other input than the data set itself and some randomness.
	
	In this experiment, the input data set is comprised of the trained weights and biases of the shadow models. We apply this to the three data sets MNIST, Census and UTKFace.
	As before, we have used \num{2000} shadow models -- \num{1000} models trained with a property \A training data set and \num{1000} models with property \B accordingly.
	Our goal of this experiment is to find out to which extend the trained weights and biases are influenced by a statistical property of the training data set.
	In particular, if the data agnostic approach t-SNE is able to cluster models with different properties apart, we can assume the influence of a property on the weights and biases of a ML model to be significant.
	
	\begin{figure}%[h!b]
		\centering
		\includegraphics[width=0.32\linewidth, trim=15 5 5 5, clip]{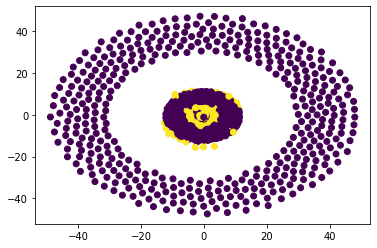}
		\includegraphics[width=0.32\linewidth, trim=15 5 5 5, clip]{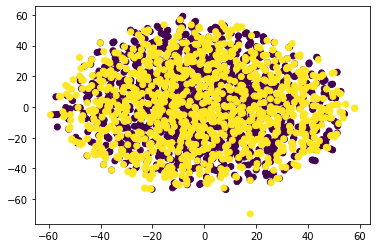}
		\includegraphics[width=0.32\linewidth, trim=15 5 5 5, clip]{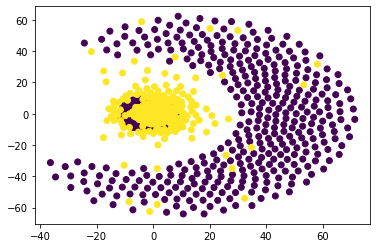}
		\caption{t-SNE visualization of MNIST (left), Census (middle) and UTKFace (right) models. Each yellow dot is a model with property~$\mathbb{A}$, each purple dots represents a model with property~$\mathbb{B}$ (\num{1000} models per property and data set).}
		\label{fig:tsne}
	\end{figure}
	
	\subsection*{t-SNE results}
	As depicted in Figure~\ref{fig:tsne}, t-SNE has produced a well defined clustering for the two image data sets MNIST and UTKFace:
	models trained with property \A training data sets (yellow dots) are placed close to the center of the visualization, while property \B models (purple dots) are mostly further from the center.
	This indicates that the properties, defined in Section~\ref{sec:pia-adv} for MNIST and UTKFace, heavily influence the weights and biases of the trained models.
	In fact, without any additional information about the parameters or the properties of the underlying training data sets, t-SNE is able to distinguish the models by property with surprisingly high accuracy.
	Based on these results, one could construct a simple PI adversary $\mathcal{A}^{\textit{t-SNE}}$ by measuring the euclidean distance $\ell$ of a target model from the center of the t-SNE clustering.
	If $\ell$ is below a certain threshold for a target model~\targetM, $\mathcal{A}^{\textit{t-SNE}}$ infers property~\A, otherwise it infers property~\B.
	For MNIST, $\mathcal{A}^{\textit{t-SNE}}$ has 86.7\% accuracy based on our experiment, while the UTKFace $\mathcal{A}^{\textit{t-SNE}}$ has 72.0\% accuracy.
	We stress that these two $\mathcal{A}^{\textit{t-SNE}}$ are solely based on the t-SNE visualization of the model parameters, no training on shadow models is needed.
	
	However for Census, t-SNE has not clustered models with different properties of their training data sets apart (see second visualization in Figure~\ref{fig:tsne}).
	In contrast to the other two data sets MNIST and UTKFace, Census is a tabular data set.
	It also may be that the properties defined in Section~\ref{sec:pia-adv} have a smaller imminent impact on the weights and biases during training.
	We leave a more profound analysis of possible reasons for the different behavior of the t-SNE visualization on the three data sets for future work.
	
	\section{Experiment 5: Preprocessing of training data}
	\label{sec:preprocessing}
	In addition to the experiments above, we have analyzed the impact of simple training data preprocessing techniques on the success of PI attacks. 
	For the two image data sets MNIST and UTKFace, we have added two different versions of noise on the images of the training data set: Gaussian noise and ``salt-and-pepper'' (SNP) noise -- the latter being sparsely occurring white and black pixels added on an image.
	For UTKFace, we have also tested mirroring the training images horizontally\footnote{Mirroring the images of the training data set has not been performed on the MNIST data set, since mirroring images of digits does not always produce valid digits. In contrast, the visual features of human faces in the UTKFace data set are generally symmetric.}.
	
	For the tabular data set Census, we have tested k-anonymization~\cite{sweeney2002k}, censoring the PIA target attribute and training on a synthetic data set (based on the original Census data set).
	
	\subsection*{Preprocessing results}
\begin{figure}
		\includegraphics[width=0.99\linewidth]{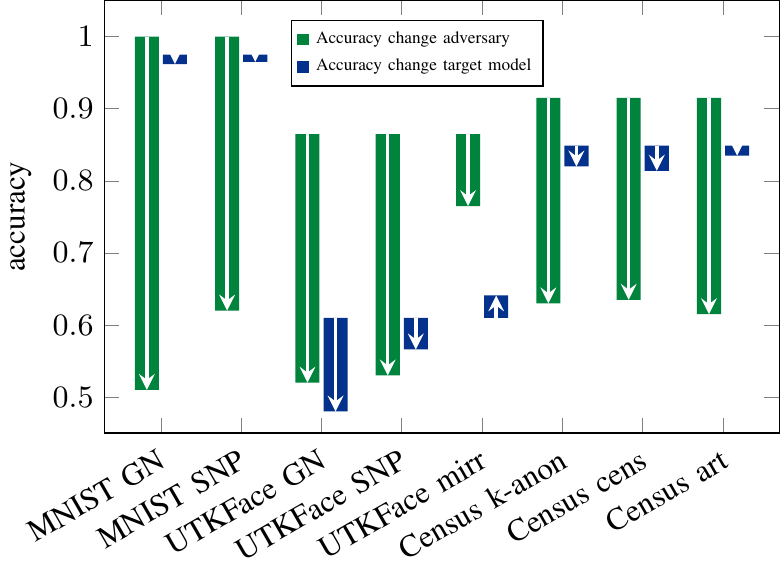}
		\caption{The results for different approaches of preprocessing techniques for the three data sets MNIST, Census and UTKFace: The mean accuracy loss of the attacker is depicted in green (defense efficiency) and the mean accuracy loss of the target models regarding their task is shown in blue (model utility loss). GN=Gaussian noise, SNP=salt and pepper noise, mirr=mirroring images, k-anon=k-anonymization, cens=censoring of PIA target attribute, art=training on artificial dataset (based on original Census).}
		\label{fig:preprocessing}
	\end{figure}

	We now discuss the most relevant results of the preprocessing experiment here and point the interested reader to Appendix~\ref{appendix-preprocessing} for a more detailed description of the experiments. 
	A summary of our results is depicted in Figure~\ref{fig:preprocessing}.
	
	\paragraph{MNIST} 
	For the MNIST data set, noising the images of the training data set is very effective:
	This drops the PIA accuracy from 100\% to 51\% by applying Gaussian noise, and to 62\% by applying SNP noise.
	Since the task of the PI adversary \advA is binary, this is very close to the 50\%~baseline of random guessing.
	The accuracy of the target model has not suffered significantly by noising the training data set: It has dropped by only 1.3\%P (Gaussian), respectively 1\%P for SNP.

	\paragraph{UTKFace}
	Adding noise to the images of the training data set has been similarly successful for the UTKFace data set:
	From the original PI attack accuracy of 86.5\%, noising the training data set has resulted in 52\% attack accuracy (Gaussian) and 53\% respectively (SNP).
	The accuracy of the target model has dropped more significantly than the MNIST target model, by 13\%P and 4.4\%P for Gaussian noise and SNP, respectively.
	On the UTKFace data set, we have also tested mirroring the images horizontally.
	While reducing the adversary's accuracy by only 10\%P to 76.5\%, this has \emph{improved} the accuracy of the target model by 3.1\%P.
	
	\paragraph{Census}
	All three preprocessing methods are similarly successful for the tabular data:
	The PI accuracy has dropped from originally 91.5\% to 63\% using k-anonymization, to 63.5\% with censoring the PIA target attribute, and to 61.5\% applying an artificial data set.
	Accordingly, the accuracy of the target model drops by 1.4\%P--3.5\%P.
	Regarding censoring the target property, \citeauthor{zhang2021leakage} have produced similar results, showing that this does not prevent the property to be adversarially inferred from a target model~\cite{zhang2021leakage}.
	
	\paragraph{Summary and disclaimers}
	While the results for MNIST may seem best in this experiment, we want to stress that the artificial target property in this experiment (an increase of the images' gamma value; see Appendix~\ref{appendix-preprocessing}) may not have relevant implications for practical properties, such as the race distributions of UTKFace.
	In this experiment, however, the UTKFace target models have had a low accuracy to begin with. This may be the main reason why mirroring the images even improves the accuracy.
	The most realistic preprocessing experiment is most likely the Census experiment, where creating an artificial data set has shown the best results. 
	Here, we can see a significant impact on the adversary's attack accuracy by generating artificial data (a drop of 30\%P in accuracy), but with 61.5\% accuracy left, the adversary does still have a significant advantage over random guessing (50\%).
	
	\section{Discussion}
	\label{sec:discussion}
	
	With the above experiments, we have tested different defense strategies to defend against property inference attacks and have conducted an in-depth analyses.
	This section compares and discusses our results to yield insights for future research in the yet unexplored field of defending against PIAs.
	
	\subsection{Choosing the right defense approach}
	We have introduced defense mechanisms at different stages of the machine learning pipeline. 
	Both \pu experiments are positioned after the target model training and before its prediction phase, respectively its publication. 
	In contrast, the preprocessing approach is applied prior to the training.
	Since most machine learning algorithms require several data preprocessing steps, implementing a defense mechanism based on preprocessing training data could be easily adapted in real-world scenarios.
	At least for tabular data, our experiments in Section~\ref{sec:preprocessing} have shown a good privacy-utility trade-off, especially the artificial data approach.
	
	Nevertheless, depending on the organization and application scenario of a machine learning model, a post-training approach like \pu might have its benefits as well.
	Further experiments could test the combination of both pre- and post-training approaches.
	Since both of them are not promising to provide the generic PIA defense we aimed for, we assume the combination of both does not significantly improve the defense.
	Instead, we suggest to focus further analyses on other approaches \emph{during} the training, as laid out in Section~\ref{sec:futurework}.
	
	%\todo[inline]{which approach is more applicable in real-world scenario?}
	%\todo[inline]{which approach is easier to adapt after new samples are added to the data set/was further trained with new data points?}
	%\todo[inline]{future work: use preprocessing and property unlearning defenses together}

	%TIME COSTS (Rest was moved to the PU section in main text)
	%Our preprocessing experiments (see Section~\ref{sec:preprocessing}) have a low footprint since we have focused on testing \emph{simple} preprocessing methods.
	%The most advanced method was generating artificial data for the Census data set, which has not taken more than a minute of computing time on a modern desktop computer.

	\subsection{Measuring defense effectiveness}
	%\todo[inline]{JW: are there any facts or metrics that can measure the protection level? or the effectiveness of the different approaches? How could you compare different defense mechanisms? - random ideas: a) use lime to visually compare and build a distance metric, b) use the same cross-validation procedure for the comparison that you used to find out that PU is not that good and then build a metric how many models are covered with the defense or so... }
	
	We have analyzed the effectiveness of our defense mechanisms based on their privacy-utility trade-offs, i.e., the accuracy decrease of the adversary (privacy) versus the corresponding loss in accuracy of the target models regarding their trained task (utility).
	This is practical and reasonable within one study to demonstrate and compare the impact of a particular defense strategy on a certain target model for a given data set.
	However, we see the need for a metric to compare the impact of two different PIA defense mechanisms, which is independent of specific target models but instantiated for a dedicated training data set.
	This would enable the research community to compare varying defense mechanisms across different works.
	One approach is formulated by \citeauthor{suri2021formalizing}~\cite{suri2021formalizing} suggesting to model the adversary's knowledge of the underlying distribution.
	Their comparison is quantified by measuring the observed leakage through a PIA in contrast to the information leakage if random samples from the actual training data set were directly leaked. 
	While this approach is rather theoretical in nature, a more practical approach is \emph{ML-Doctor} by \citeauthor{liu2021ml}~\cite{liu2021ml}.
	The authors have published a software toolkit\footnote{\url{https://github.com/liuyugeng/ML-Doctor}} to compare the leakage of machine learning models for four different privacy attacks -- unfortunately not including PIAs.
	An extension of ML-Doctor to take PIAs into account would be desirable.

	\subsection{Lessons learned}
	\label{sec:lessons-learned}
	With our cross-validation experiment in Section~\ref{sec:exp2-results}, we have shown how PI adversaries react to \textbf{\pu} in different ways.
	Specifically, some adversaries could still reliably infer training data properties after 15 \pu iterations, while other adversaries reliably inferred the \emph{wrong} property after the same process.
	This shows that it is \textbf{hard to utilize a post-training technique} like \pu as a generic defense against a whole class of PI adversaries:
	On the one hand, one needs to defend against the strongest possible adversaries while on the other hand, it is important not to introduce additional leakage by adapting the target model too much.
	Depending on the adversary instance (of which we have tested 20), most of our target models clearly show one of these deficiencies after 15 rounds of \pu (see Section~\ref{sec:exp2-results}).
	
	Our t-SNE experiment in Section~\ref{sec:t-sne} shows that at least for image data sets, statistical \textbf{properties of training data sets have a severe impact} on the trained parameters of a machine learning model.
	This is in line with the experiment of Section~\ref{sec:lime}, which shows how two PI adversaries with the same objective focus on different parts of the parameters of a target model.
	If a property is manifested in many areas of a model's parameters, PI adversaries can rely on different regions.
	This implies that completely pruning such properties from a target model after training is hard or even impossible, without significantly harming its utility.
	
	This brings us to the next experiment involving different \textbf{preprocessing} methods (Section~\ref{sec:preprocessing}).
	Different from techniques like differential privacy, we have deliberately chosen to test the context-independent mechanisms such as mirroring and adding noise to image data sets, as well as k-anonymizing, censoring and remodeling tabular data as an artificial data set.
	Our results show that such approaches do bear potential.
	Although they mostly imply a drop of accuracy in the target models, the observed performance loss was small for many of the tested methods.
	We believe artificially generated data to be the most promising technique and suggest further experiments in the next section.	
	
	\subsection{Future work}
	\label{sec:futurework}
	
	\paragraph{Preprocessing training data}
	We have not tested training data preprocessing in an \emph{adaptive} environment yet, where the adversary would adapt to the preprocessing steps and retrain on shadow models with preprocessed training data as well.
	Intuitively, this would weaken the defense while costing the same utility in the target models.
	Still, experiments comparing different techniques in an adaptive setting could aid the community in pointing to promising directions for future research.
	
	Furthermore, as the technique with most potential for defending against PIAs for tabular data, the generation of artificial data could be further explored:
	One could adapt the synthesis algorithm such that statistical properties are arbitrarily modified in the generated data set.
	A similar goal is pursued in many bias prevention approaches in the area of fair machine learning, e.g., in~\cite{sharma2020data}.
	
	\paragraph{Adapting the training process}
	Another method from a similar area called \emph{fair representation learning} is punishing the model when learning biased information by introducing a regularization term in the loss function during training~\cite{buyl2020debayes, zemel2013learning, creager2019flexibly}.
	As a defense strategy against PIAs, one would need to introduce a loss term which expresses the current property manifestation within the model and causes the model to hide this information as good as possible.
	In theory, this would be a very efficient way to prevent the property from being embedded in the model parameters.
	Since it would be incorporated into the training process, the side effects on the utility of the target model should be low.
	
	\paragraph{Post-training methods}
	\citeauthor{liu2021ml}~\cite{liu2021ml} experiment with knowledge distillation~(KD) as a defense mechanism against different privacy attacks like membership inference.
	The idea is to decrease the number of neurons in a neural network in order to lower its memory capacity.
	Unfortunately, the authors do not consider PIAs -- it would be interesting to see the impact of KD on its success rate.
		
	\section{Conclusion}
	\label{sec:conclusion}
	In this paper, we performed the first extensive analysis on different defense strategies against white-box property inference attacks.
	This analysis includes a series of thorough experiments on \emph{\pu}, a novel approach which we have developed as a dedicated PIA defense mechanism.
	Our experiments show the strengths of \pu when defending against a dedicated adversary instance.
	But they also highlight its limits, in particular its lacking ability to generalize.
	We elaborated on the reasons of this limitation and concluded with the conjecture that statistical properties of training data are deep-seated in the trained parameters of machine learning models.
	This allows property inference adversaries to focus on different parts of the parameters when inferring such properties, but also opens up possibilities for much simpler attacks, as we have shown via t-SNE model parameter visualizations.
	
	Apart from the \emph{post-training} defense \pu, we have also tested different training data \emph{preprocessing} methods.
	Although most of them were not directly targeted at the sensitive property of the training data, some of the methods have shown promising results.
	In particular, we believe that deliberately generating a property-free, artificial data set based on the distribution of an original training data set could be a candidate for a PIA defense with very good privacy-utility tradeoff.
	We conclude with many promising directions for future research, such as adapting the training process by introducing additional loss terms or applying knowledge distillation to target models.\\
	
	\paragraph{Acknowledgements}
	We would like to thank Alexander Klassen for performing the preprocessing experiment (Section~\ref{sec:preprocessing} and Appendix~\ref{appendix-preprocessing}).
	We also wish to thank Anshuman Suri for valuable discussions and we are grateful to the anonymous reviewers of previous versions for their feedback.

	\bibliography{property-unlearning-paper}
	
	\newpage
	\appendices
	
	\section{Multi-property unlearning}
	\label{appendix-multi-prop}
	
	\subsection{Data set construction}
	\label{subsec:multi-prop-setup}

	\begin{table*}[t]
		\centering
		\caption{The data sets used for the experiments. avg=average, distrib.=distribution, PIA=Property Inference Attack, Acc. = Accuracy, \mbox{PU=Property Unlearning.}}
		\begin{tabular}{ccccccc}
			\toprule
			\multirow{3}{*}{\textbf{Experiment}} &
			%\multirow{2}{*}{\textbf{Data Set}} &
			\multirow{3}{*}{\textbf{Size}} &
			\multirow{3}{1.5cm}{\centering \textbf{Task}} &
			\multirow{3}{*}{\textbf{Target Property}} &
			\multirow{3}{*}{\textbf{$|DS_{*}|$}} &
			\multirow{3}{1.5cm}{\centering \textbf{Initial PIA Accuracy}} &
			\multirow{3}{2.5cm}{\centering \textbf{Avg. Task Acc. Loss due to PU}} \\
			\addlinespace[18pt]
			%\midrule
			\midrule
						\expMultiAR, \expMultiRA %& UTKFace (multi)
						& 23K & gender & race / age distrib. & 10K & 91--98\% & 0.9--1.7\%P \\
			\bottomrule
		\end{tabular}
		\label{table:dataset-multi}
	\end{table*}

	Apart from the single property experiments as described in Section~\ref{sec:pu-exp}, we have constructed one data set with \emph{two} sets of properties based on the UTKFace data set.
	In addition to the property \emph{race} as described above, we introduce another property \emph{age}:
	$\mathbb{A}_\text{UTK}^\text{age}$ contains instances with an age larger than 25, most of them having an age larger than 39.
	In contrast, $\mathbb{B}_\text{UTK}^\text{age}$ contains instances with an age lower than 40, most of them younger than 26.
	Note that the two data sets $DS_{\mathbb{A}_\text{UTK}^\text{age}}$ and $DS_{\mathbb{B}_\text{UTK}^\text{age}}$ slightly overlap.
	We have opted for this small overlap in order to ensure that both data sets $DS_\mathbb{A}^\text{multi}$ and $DS_\mathbb{B}^\text{multi}$ (as described below) contain \num{10000} images.
	Training the models with less training data could have led to significantly less initial utility of our shadow and test models.
	
	For the multi-property unlearning experiment, we use two test data sets:
	$DS_\mathbb{A}^\text{multi}$ only consisting of instances with both properties $\mathbb{A}_\text{UTK}^\text{race}$ and $\mathbb{A}_\text{UTK}^\text{age}$ (``old white''), and its counterpart $DS_\mathbb{B}^\text{multi}$ only containing instances with both properties $\mathbb{B}_\text{UTK}^\text{race}$ and $\mathbb{B}_\text{UTK}^\text{age}$ (``young non-white'').
	The models for these two data sets have been constructed and trained in the same way as the UTKFace models described in the previous section.
	
	While the models in the \expUtk experiment of Section~\ref{sec:pu-exp} are each trained on images of the whole range of ages, the auxiliary data sets in the multi-property unlearning experiments are either trained on ``old white'' or ``young non-white'' data sets.
	With these two properties combined, they are trained on rather specialized data sets, opposed to the models in the \expUtk, which are trained on more diverse data sets (only divided by the property \emph{age} and not by the property \emph{race}).
	Since the test data set is composed of images of \emph{all} age and race ranges, the test accuracy is slightly lower at 86.6--85.8\% for the models in the multi-property experiments.
	
	We unlearn the properties consecutively, i.e., the property of the categories \emph{age} and \emph{race} are unlearned one after the other.
	This yields two sub-experiments, which we will refer to as \expMultiAR and \expMultiRA.
	For the unlearning process, we need two separate adversarial meta classifiers, one for each property category: $\mathcal{A}^\text{age}$ and $\mathcal{A}^\text{race}$.
	Because these two adversaries have been trained on models with slightly more diverse data sets (as explained above), they perform worse on the multi-property models. While $\mathcal{A}^\text{race}$ reaches a test accuracy of 91\%, $\mathcal{A}^\text{age}$ scores 98\%, in comparison to 99.8\% in the more diverse case above.
	
	\subsection{Multi-property unlearning results}
	\label{subsec:multi-prop-results}
	
	During the multi-\pu experiments, we have applied our \pu strategy consecutively on UTKFace target models with the two sets of properties \emph{age} and \emph{race}, as described above.
	No matter in which order we have unlearned the properties, the performance of our \pu approach was similar to the experiments in Section~\ref{sec:pu-exp}.
	
	\begin{figure}[h!t]
		\centering
		% \begin{subfigure}{0.48\textwidth}
		\centering
		\subfloat[\label{fig:utk2-acc-age-race-before}]{
			\includegraphics[height=\figheight]{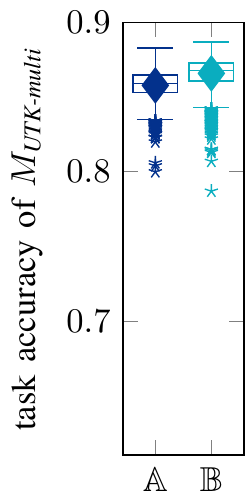}}
		\subfloat[\label{fig:utk2-acc-age-race-after_p1}]{
			\includegraphics[height=\figheight, trim=16 0 0 0,clip]{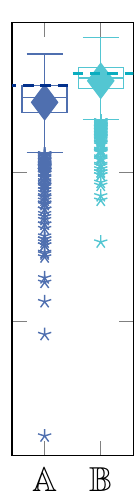}}
		\subfloat[\label{fig:utk2-acc-age-race-after_p2}]{
			\includegraphics[height=\figheight]{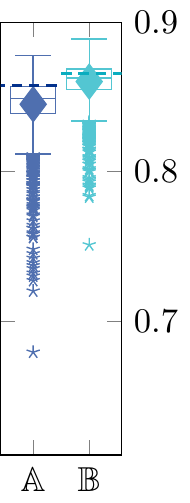}}
		\caption{Accuracy of the target models \protect\subref{fig:utk2-acc-age-race-before} before the \expMultiAR experiment, \protect\subref{fig:utk2-acc-age-race-after_p1} after unlearning the properties for \emph{age}, and \protect\subref{fig:utk2-acc-age-race-after_p2} after unlearning the \emph{race} properties.}
		\label{fig:utk2-acc-age-race}
	\end{figure}
	% \end{subfigure}
	% \hfill
	% \begin{subfigure}{0.48\textwidth}
	\begin{figure}[h!t]
		\centering
		\subfloat[\label{fig:utk2-acc-race-age-before}]{
			\includegraphics[height=\figheight]{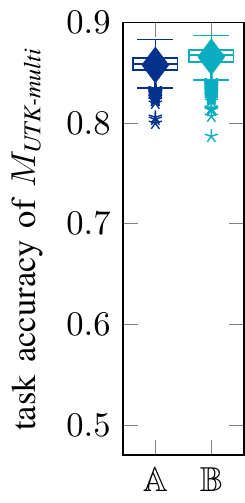}}
		\subfloat[\label{fig:utk2-acc-race-age-after_p1}]{
			\includegraphics[height=\figheight, trim=16 0 0 0,clip]{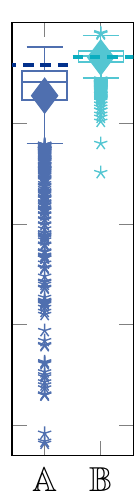}}
		\subfloat[\label{fig:utk2-acc-race-age-after_p2}]{
			\includegraphics[height=\figheight]{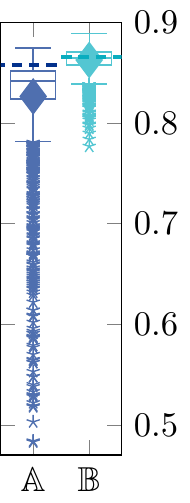}}
		\caption{Accuracy of the target models \protect\subref{fig:utk2-acc-race-age-before} before the \expMultiRA experiment, \protect\subref{fig:utk2-acc-race-age-after_p1} after unlearning the properties for \emph{race}, and \protect\subref{fig:utk2-acc-race-age-after_p2} after unlearning the \emph{age} properties.}
		\label{fig:utk2-acc-race-age}
		% \end{subfigure}
		% \caption{Accuracy Experiments. TODO (sort out letters/numbering)}
	\end{figure}
	
	\begin{figure}[h!t]
		\centering
		% \begin{subfigure}{0.48\textwidth}
		\centering
		\subfloat[\label{fig:utk2-eff-age-before}]{
			\includegraphics[height=\figheight]{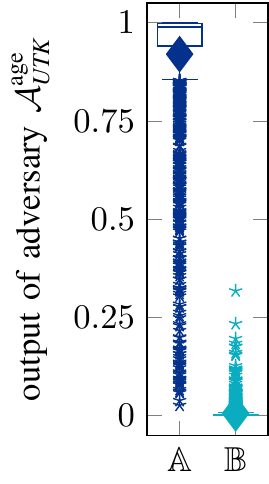}}
		\subfloat[\label{fig:utk2-eff-age-after-a}]{
			\includegraphics[height=\figheight, trim=24 0 0 0,clip]{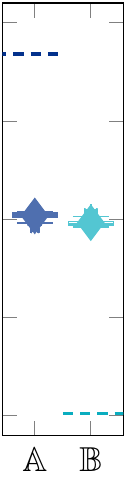}}
		\subfloat[\label{fig:utk2-eff-age-after-r}]{
			\includegraphics[height=\figheight, trim=24 0 0 0,clip]{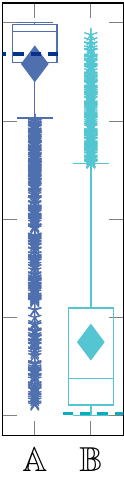}}
		\subfloat[\label{fig:utk2-eff-age-after-ra}]{
			\includegraphics[height=\figheight, trim=0 0 24 0,clip]{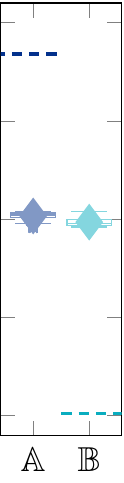}}
		\caption{The efficiency of \pu in our multi-\pu experiments regarding the two properties for \emph{age}. \protect\subref{fig:utk2-eff-age-before} baseline before the experiments, \protect\subref{fig:utk2-eff-age-after-a} after unlearning the properties for \emph{age} in the \expMultiAR experiment, \protect\subref{fig:utk2-eff-age-after-r} after unlearning the \emph{race} properties in the \expMultiRA experiment, \protect\subref{fig:utk2-eff-age-after-ra} after unlearning the properties for \emph{race} and \emph{age} consecutively in the \expMultiRA experiment.}
		\label{fig:utk2-eff-age}
	\end{figure}
	% \end{subfigure}
	% \hfill
	% \begin{subfigure}{0.48\textwidth}
	\begin{figure}[h!t]
		\centering
		\subfloat[\label{fig:utk2-eff-race-before}]{
			\includegraphics[height=\figheight]{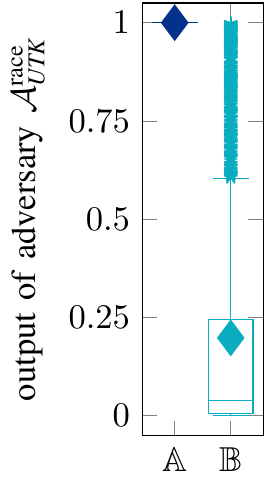}}
		\subfloat[\label{fig:utk2-eff-race-after-r}]{
			\includegraphics[height=\figheight, trim=24 0 0 0,clip]{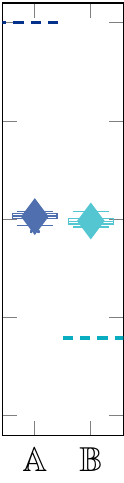}}
		\subfloat[\label{fig:utk2-eff-race-after-a}]{
			\includegraphics[height=\figheight, trim=24 0 0 0,clip]{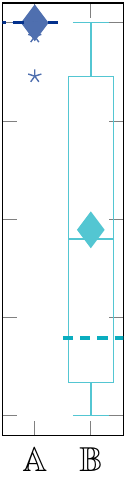}}
		\subfloat[\label{fig:utk2-eff-race-after-ar}]{
			\includegraphics[height=\figheight, trim=0 0 24 0,clip]{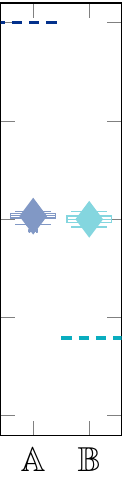}}
		\caption{The efficiency of \pu in our multi-\pu experiments regarding the two properties for \emph{race}. \protect\subref{fig:utk2-eff-race-before} baseline before the experiments, \protect\subref{fig:utk2-eff-race-after-r} after unlearning the properties for \emph{age} in the \expMultiAR experiment, \protect\subref{fig:utk2-eff-race-after-a} after unlearning the \emph{race} properties in the \expMultiRA experiment, \protect\subref{fig:utk2-eff-race-after-ar} after unlearning the properties for \emph{race} and \emph{age} consecutively in the \expMultiAR experiment.}
		\label{fig:utk2-eff-race}
		% \end{subfigure}
		% \caption{PU Efficiency Experiments. TODO (sort out letters/numbering)}
	\end{figure}
	
	We begin by presenting the results of the \expMultiRA experiment, i.e., unlearning the properties $\mathbb{A}_\text{UTK}^\text{race}$ and $\mathbb{B}_\text{UTK}^\text{race}$ first and then unlearning the properties $\mathbb{A}_\text{UTK}^\text{age}$ and $\mathbb{B}_\text{UTK}^\text{age}$ afterwards.
	As mentioned above in Section~\ref{subsec:multi-prop-setup}, the adversaries do not perform as well as in the previous experiments, because they are not tailored to the target models in these experiments with a less diverse training set (see Figures~\ref{fig:utk2-eff-age} and~\ref{fig:utk2-eff-race}).
	This is apparent in the adversary output for $\mathbb{A}^\text{age}$ and $\mathbb{B}^\text{race}$ before unlearning~(see Figures~\ref{fig:utk2-eff-age-before} and~\ref{fig:utk2-eff-race-before}), where some instances are misclassified, although the average output is still reasonably good at $0.9$ and $0.2$, respectively.
	
	Interestingly, after unlearning the \emph{other} property~$\mathbb{B}^\text{age}$ from the target models~$\mathcal{M}$, the performance of adversary $\mathcal{A}^\text{race}_\textit{UTK}$ has already decreased significantly regarding property~$\mathbb{B}^\text{race}$:
	$\mathcal{A}^\text{race}_\textit{UTK}$ has an average output of $0.47$ for property~$\mathbb{B}^\text{race}$ in this case, i.e., many of target models are close to the \pu goal regarding $\mathbb{B}^\text{race}$ already (see Figure~\ref{fig:utk2-eff-race-after-a}).
	However, the observed impact of unlearning the \emph{other} property has not been as big as in the other cases, see property~$\mathbb{A}^\text{race}$ in Figure~\ref{fig:utk2-eff-race-after-a} or both properties in Figure~\ref{fig:utk2-eff-age-after-r}.

	As our results in Figures~\ref{fig:utk2-eff-age-after-ra}, \ref{fig:utk2-eff-age-after-a}, \ref{fig:utk2-eff-race-after-r} and \ref{fig:utk2-eff-race-after-ar} show, all target models in the \expMultiAR and \expMultiRA experiments could be hardened successfully via \pu.
	
	Referring to Figure~\ref{fig:utk2-acc-age-race} of the \expMultiAR experiment, the accuracy of the target models does not drop significantly more than in the previous experiments due to \pu.
	For some models, the drop is significant (momentarily down to $62\%$ accuracy), but the average drop is at 1.2\%P for models trained with properties~\A\ ($\mathbb{A}^\text{age}$ and $\mathbb{A}^\text{race}$) and 0.5\%P for the target models trained with properties~\B.
	Across both property sets, this yields an average accuracy loss of 0.9\%P for the \expMultiAR experiment.
	
	The results for the \expMultiRA experiment are similar, as depicted in Figure~\ref{fig:utk2-acc-race-age}, although there is more significant task accuracy loss for the target models during the unlearning of the property~$\mathbb{A}^\text{race}$.
	This is due to the very good performance of the adversary $\mathcal{A}^\text{race}_\textit{UTK}$ against the original target models for property~\A (see Figure~\ref{fig:utk2-eff-race-before}).
	For 293 of the \num{2000} target model instances, $\mathcal{A}^\text{race}_\textit{UTK}$ initially had perfect confidence for the $\mathbb{A}^\text{race}$ property, i.e., output of $1.0$ was returned.
	Some random modifications need to be applied to these target models' parameters to initiate the \pu process.
	Random parameter flips in combination with the usual \pu steps finally led to a slightly larger loss, on average 3.1\%P for models with the properties \A ($\mathbb{A}^\text{age}$ and $\mathbb{A}^\text{race}$), and 0.5\%P for models trained with properties \B.
	In total, we measured an accuracy loss of 1.7\%P across both property sets in the \expMultiRA experiment. %The additional measures will be further discussed in Section~\ref{subsec:perfect_confidence}.
	
	\section{Details on preprocessing experiments}
	\label{appendix-preprocessing}
	
	\begin{table}[h!t]
		\centering
		\caption{The data sets used for preprocessing experiments. avg=average, PIA=Property Inference Attack, Acc. = Accuracy, \mbox{PU=Property Unlearning.}}
		\begin{tabular}{ccccc}
			\toprule
			\multirow{3}{*}{\textbf{Exp.}} &
			\multirow{3}{*}{\centering \textbf{Task}} &
			\multirow{3}{*}{\textbf{Task Acc.}} &
			\multirow{3}{*}{\textbf{Target Property}} &
			\multirow{3}{*}{\textbf{PIA Acc.}} \\
			\addlinespace[18pt]
			\midrule
			\expMnistPre & digits &  97.5\% & gamma & 100\% \\
			\midrule
			\expCensusPre & income & 84.9 & gender distrib. & 91.5\%  \\
			\midrule
			\expUtkPre & age & 61.0 & race distrib. & 86.5\% \\
			\bottomrule
		\end{tabular}
		\label{table:prepro_data_sets}
	\end{table}
	
	For the preprocessing experiments, we have used different configurations of the PI attack, see Table~\ref{table:prepro_data_sets}.
	Specifically, the target property for MNIST is whether the training images have been darkened by a gamma-value of 5 (instead of the property \emph{noise} in \expMnist before).
	Also, the task for UTKFace has been changed to age recognition (with a worse performance than in the previous experiment \expUtk) and the target property is whether images with the class \emph{black} is present in the training data set (in the original data set, 19\% of images have this class) or these images have been removed from the database.
	For all data sets, we have plotted and discussed the average result of 40 experiments (=40 target models).
	
	\paragraph{Noise parameters}
	The best privacy-utility trade-off we have observed for Gaussian noise on the image data sets is with a variance of 0.1. 
	For ``salt and pepper'', we have chosen a 10\% of noisy pixels for the same reason.

	\paragraph{k-Anonymization}
	We have k-anonymized the Census data set with an implementation of the Mondrian algorithm~\cite{lefevre2006mondrian} and we have chosen the following attributes as \emph{quasi-identifiers}: age, race, sex, marital-status and relationship.
	
	\paragraph{Censoring}
	To adapt the tabular Census data set for ANN usage, we encoded all textual attributes into numbers.
	E.g., the attribute \emph{gender} has been mapped from $\{\textit{male}, \textit{female}\}$ to $\{0,1\}$.
	We have censored the PIA target property \emph{gender} by replacing all zeros and ones of the gender attribute by 0.5.
	
	\paragraph{Artificial data}
	To generate artificial data from the Census data set, we have used the open source implementation of \emph{CTGAN}~\cite{xu2019modeling}.
	We have not modified the generation of the artificial data set, i.e., we have not taken any extra steps to hide the property in the generated data set.
	
%	\section{Perfect adversarial confidence}
%	\label{subsec:perfect_confidence}
%	Calculating these gradients is done automatically by the TensorFlow  framework~\cite{tensorflow}.
%	However, if the adversary has perfect confidence that a target model's training data set has some property, i.e., the output for this property is 1, gradients cannot be calculated.
%	To circumvent the problem, we randomly flip a low number of the target model's weights, multiplying them by -1, until the adversary is less confident. % (see lines~\ref{algline:while-perf-start}--\ref{algline:while-perf-end}).
	
\end{document}